\documentstyle[12pt,axodraw,epsf]{article} \voffset0cm \hoffset0cm
\begin{document}

\begin{flushright}
CERN-TH/99-230\\[-0.15cm]
MPI-Th/99-30\\[-0.15cm]
FERMILAB-PUB-99/215-T\\[-0.15cm]
THES-TP/99-09
\end{flushright}
\vspace{-3.cm}
\begin{flushleft}
hep-ph/9907522\\
July 1999
\end{flushleft}
\vspace{0.5cm}

\begin{center}
{\Large {\bf Cumulative Non-Decoupling Effects of Kaluza--Klein}}\\[0.2cm]
{\Large {\bf Neutrinos in Electroweak Processes}}\\[1.5cm]
  {\large Ara Ioannisian$^a$\footnote[1]{On leave of absence from Yerevan
    Physics Institute, Alikhanian Br.\ 2, 375036 Yerevan, Armenia}
    and Apostolos Pilaftsis$^{b,c,d}$}\\[0.35cm]
  $^a${\em Max-Planck-Institut f\"ur Physik, F\"ohringer Ring 6, 80805
    Munich, Germany}\\[0.2cm]
  $^b${\em Theory Division, CERN, CH-1211 Geneva 23, Switzerland}\\[0.2cm]
  $^c${\em Fermilab, P.O. Box 500, Batavia IL 60510, U.S.A.}\\[0.2cm]
  $^d${\em Department of Theoretical Physics, University of Thessaloniki,}\\
  {\em GR 54006 Thessaloniki, Greece}
\end{center}
\vskip1.cm   \centerline{\bf  ABSTRACT}  
In Kaluza--Klein theories of  low-scale quantum gravity, gravitons and
isosinglet neutrinos may propagate in  a higher-dimensional space with
large compact dimensions, whereas all particles  of the Standard Model
are   confined      on   a   ($1+3$)-dimensional    subspace.    After
compactification   of   the extra  dimensions,  the  resulting  Yukawa
couplings of the Kaluza--Klein neutrinos to the lepton doublets become
naturally very suppressed by   a higher-dimensional volume factor,  in
agreement with phenomenological  observations.  We show that  one-loop
effects induced  by Kaluza--Klein neutrinos, albeit tiny individually,
act  cumulatively in  electroweak    processes,   giving rise  to    a
non-decoupling behaviour for  large values  of the  higher-dimensional
Yukawa  couplings.     Owing  to    the  non-decoupling   effects   of
Kaluza--Klein neutrinos, we  can  derive stronger constraints  on  the
parameters of the theory  that  originate from the non-observation  of
flavour-violating   and universality-breaking phenomena, which involve
the $W$ and $Z$ bosons, and the $e$, $\mu$ and $\tau$ leptons.\\[0.2cm]
PACS no.: 11.25.Mj

\newpage

\setcounter{equation}{0}
\section{Introduction}
Recently,  Arkani-Hamed,    Dimopoulos   and Dvali   \cite{ADD}   have
considered the  radical   possibility that the  fundamental  scale  of
quantum gravity is no   longer set by the   Planck mass $M_{\rm  P}  =
1.2\times 10^{19}$ GeV, but the  true scale of quantum gravity, $M_F$,
is many orders of magnitude smaller than $M_{\rm P}$  close to the TeV
energies.\footnote{There  have   been  earlier considerations  in  the
  literature that discussed the possibility of lowering the string but
  not the   Planck  scale in superstring  theories.   Most noticeably,
  Witten  \cite{Witten},  and    Ho$\check{{\rm   r}}$ava and   Witten
  \cite{HW} considered a  novel $M$-theory scenario compactified on an
  orbifold  $S^1/Z_2$, in which  the  string scale  was lowered by two
  orders of magnitude to energies of order $10^{16}$ GeV.  Along these
  lines,  Lykken \cite{JL}  discussed an analogous  scenario, in which
  the string scale  was further lowered to the  TeV range.  In a  much
  earlier work,   Antoniadis  \cite{IA} discussed  the possibility  of
  lowering the  compactification radius of  gauge  interactions at the
  TeV  scale  in  the    context of   string  theories.   In   related
  supersymmetric scenarios,  Dienes,  Dudas and Gherghetta \cite{DDG0}
  have   recently     studied  several  aspects    of   gauge-coupling
  unification.}  The  observed weakness of  gravity is then due to the
presence of a number $\delta$ of large  extra dimensions, within which
only gravity   can  propagate  and,  most  probably,   fields that are
singlets under the Standard Model (SM) gauge group, such as isosinglet
neutrinos \cite{ADDM,DDG,AP}.   In this theoretical  framework, the SM
particles do not  feel the presence  of the extra dimensions, but  are
rather   confined to   a  $(1+3)$-dimensional  Minkowski subspace that
constitutes our observable   world.  Most interestingly, the  ordinary
Planck mass $M_{\rm P}$ is related  to the genuinely fundamental scale
$M_F$ through
\begin{equation}
  \label{Gauss}
M_{\rm P}\ \approx\ M_F\, (R\, M_F)^{\delta/2}\, ,      
\end{equation}
where $R$  denotes  the compactification radii  of the   extra compact
dimensions, which are all  taken to  be of  equal size. The  scenario,
with $\delta=1$ and $M_F$ of order TeV, predicts a visible macroscopic
compactification radius   and  is therefore  ruled out.   Many  recent
astrophysical \cite{ADD1,astro,ADDM,DDG,AP,DS}  and   phenomenological
\cite{pheno}  studies have been  devoted  to explore the viability  of
low-scale theories of quantum gravity, with $\delta \ge 2$.

In this  paper, we shall  study  the phenomenological  consequences of
loop effects of higher-dimensional  isosinglet neutrinos on  collider
and lower energy experiments.  Specifically, we  find that both at the
tree and quantum  levels, higher-dimensional isosinglet neutrinos can
naturally induce observable phenomena  of lepton-flavour violation and
universality    breaking in  $W$-   and  $Z$-boson  interactions.   To
quantitatively analyze the   new-physics effects, we shall consider  a
minimal higher-dimensional scenario, in which the SM is extended by an
isosinglet  neutrino  $N$  that  propagates   in $3+\delta$   spatial
dimensions.  The isosinglet neutrino  $N$ couples, with non-universal
Yukawa couplings, to all three  lepton SM doublets, $L_e$, $L_\mu$ and
$L_\tau$,  which are localized on   our $3$-dimensional world.   After
compactification, the  resulting Yukawa couplings  of the Kaluza--Klein
(KK) neutrinos to the SM  leptons come out to  be highly suppressed by
the  volume factor of  the    extra  dimensions $M_F/M_{\rm P}    \sim
10^{-16}$ \cite{ADDM,DDG}.

One might now think that the new-physics  phenomena mediated by the KK
neutrinos would also be extremely suppressed by the same volume factor
of  the extra dimensions.  However, this  is  not true.  After summing
over the tower  of  the KK neutrinos,   we obtain an effective  theory
whose Yukawa interactions are mediated by order-unity Yukawa couplings
of the original Lagrangian  before compactification. As a consequence,
we expect  a cumulative non-decoupling phenomenon  of the KK neutrinos
to  occur  in loops for    large higher-dimensional Yukawa  couplings,
namely the KK neutrinos appear to violate the known decoupling theorem
due to  Appelquist and  Carazzone \cite{AC}.\footnote{This  theorem is
  not directly  applicable to  spontaneous-symmetry-breaking theories,
  such as the one we are considering here.  The reason is that not all
  operators of dimension 2 can  be increased independently of those of
  dimension 3, since they  are related by  the Higgs  mechanism.}  The
higher-dimensional non-decoupling phenomenon  is analogous to the  one
studied   in \cite{KPS,BKPS,MPLAP,IP,BP,nondec}, for  singlet-neutrino
scenarios   \cite{WW,BSVMV} with large    SU(2)$_L$  Dirac masses  and
mixings.  Because   of   the non-decoupling   effects of   heavy  (KK)
neutrinos, phenomena of new physics can be dramatically enhanced to an
observable level, such  as lepton-flavour-violating decays of the  $Z$
boson    \cite{KPS}, universality-breaking   effects  in the  diagonal
leptonic decays of  the $Z$ boson  \cite{BKPS}, neutrinoless two-  and
three-body decays of the $\tau$ and $\mu$ leptons \cite{MPLAP,IP}, and
universality-breaking effects  in leptonic asymmetries measured on the
$Z$ pole   \cite{BP}.  In  fact,  the   non-observation of  the  above
new-physics effects places stringent bounds  on the parameter space of
the   theory.  Here,  we shall   perform  an  analogous study  for the
higher-dimensional singlet-neutrino scenario under consideration.  The
limits obtained  by  our analysis  are rather  generic  and can easily
carry over to related higher-dimensional models.

Another   important feature  of   the  singlet-neutrino models  is the
decoupling  property  of a   very  high isosinglet mass  \cite{SS}. In
higher-dimensional models, the  fundamental Planck mass, $M_F$, is the
one that is now playing the role of the  isosinglet mass scale.  Thus,
we  expect that the  KK neutrinos  decouple  from the loops as $M_F\to
\infty$. In this limit, all new-physics phenomena mentioned above will
be suppressed by  inverse powers  of  $M_F$.  However, for  relatively
small values of $M_F$, e.g.\ $M_F  \stackrel{<}{{}_\sim} 100$ TeV, the
screening effect of the  higher-dimensional Planck scale will  be less
dramatic, and experimental information is then needed to place a lower
bound on $M_F$.

The organization of the paper is as follows: in Section 2, we describe
the  basic   low-energy  structure    of a minimal   model   with  one
higher-dimensional  isosinglet neutrino.    In Section  3,  we derive
constraints on the parameters    of the KK  theory, which   arise from
tree-level contributions to electroweak observables.  In Section 4, we
explicitly demonstrate the   cumulative  non-decoupling effect of   KK
neutrinos  in a typical flavour-changing neutral-current (FCNC) graph.
In Section 5, we present analytic results of the loop contributions of
the KK  neutrinos to electroweak  observables of new physics, and also
set new limits on the parameters of  the theory.  Section 6 summarizes
our conclusions.

\setcounter{equation}{0}
\section{Higher-dimensional model with one singlet neutrino}

For our phenomenological study, we  shall adopt a variant \cite{AP} of
the model discussed in  Ref.\ \cite{ADDM}.  Nevertheless, the  results
of our analysis  will equally  well  apply to  other recently proposed
scenarios \cite{DDG}.  For  definiteness, we   will be considering   a
model that minimally extends the SM-field content by one singlet Dirac
neutrino, $N(x,y)$, which propagates in a $[1+(3+\delta)]$-dimensional
Minkowski space.  We denote by $x^\mu$, with $\mu  = 0,1,2,3$, the one
time and the three spatial coordinates  of our observable world and by
$y^k$,    with    $k  =   1,\dots,\delta$,   the   new   large compact
dimensions. The $y$-coordinates are compactified on a circle of radius
$R$  by applying the periodic  identification: $y \equiv  y + 2\pi R$.
Furthermore,  we  consider that  the higher-dimensional Dirac neutrino
$N(x,y)$ generally has non-universal Yukawa couplings, $\bar{h}_l$, to
the  three ordinary lepton isodoublets $L_l  (x)$, with  $l = e, \mu,
\tau$.

For the  purpose    of  illustration,  we  shall  consider   that  the
higher-dimensional Dirac neutrino  $N(x,y)$ feels the presence of only
one large  compact   dimension.  Then,  our   results can   easily  be
generalized to higher dimensions.   The leptonic sector of the minimal
model consists of the following fields:
\begin{equation}
  \label{content}
L_l(x)\ =\ \left( \begin{array}{c} \nu_{lL} (x) \\ l_L (x) \end{array}
\right) ,\qquad
l_R (x),\qquad N(x,y)\ =\ \left( \begin{array}{c} \xi (x,y) \\ 
\bar{\eta} (x,y) \end{array} \right)\, ,
\end{equation}
where $\nu_{lL}$, $l_L$  and $l_R$ describe 4-dimensional Weyl spinors
of the charged leptons and  their associate left-handed neutrinos, and
$\xi$  and  $\eta$  are two-component   spinors in 5  dimensions.  The
5-dimensional gamma matrices may be represented by
\begin{equation}
  \label{gam5}
\gamma_\mu\ =\ \left( \begin{array}{cc} 0 & \bar{\sigma}_\mu \\
                \sigma_\mu & 0 \end{array} \right) \quad {\rm and}\quad
\gamma_4\ =\ \left( \begin{array}{cc} i{\bf 1}_2 & 0 \\
                                      0 & -i{\bf 1}_2 \end{array}\right),
\end{equation}
where $\sigma^\mu = ({\bf  1}_2, \vec{\sigma} )$ and $\bar{\sigma}^\mu
= ({\bf   1}_2,  -\vec{\sigma}  )$, and $\sigma_1$,     $\sigma_2$ and
$\sigma_3$ are the  usual Pauli matrices.  The effective Lagrangian of
our minimal model reads
\begin{eqnarray}
  \label{Leff}
{\cal L}_{\rm eff}  & =&  \int\limits_0^{2\pi R}\!\! dy\
 \Big[\, \bar{N} \Big( i\gamma^\mu \partial_\mu\, +\, 
 i\gamma_4 \partial_y \Big) N\ -\ m \bar{N} N\ +\ \delta (y-a)\, 
\Big( \sum_{l=e,\mu,\tau}\, \bar{h}_l L_l\tilde{\Phi} \xi\, +\, {\rm
H.c.}\,\Big)\nonumber\\
&&+\, \delta (y-a)\, {\cal L}_{\rm SM} (\Phi )\,\Big]\, ,
\end{eqnarray}
where  $\tilde{\Phi} = i\sigma_2 \Phi^*$ and  ${\cal L}_{\rm SM} (\Phi
)$   describes the SM  Lagrangian.   The dimensionful Yukawa couplings
$\bar{h}_l$ may be related to the dimensionless ones, $h_l$, through
\begin{equation}
  \label{hl}
\bar{h}_l\ =\ \frac{h_l}{(M_F)^{\delta /2}}\ ,
\end{equation}
with $\delta =1$. Without any further restriction on the parameters of
the  theory, the reduced couplings  $h_l$ are expected  to be of order
unity, as  $M_F$ is the only  available energy  scale to normalize the
dimensionful couplings  $\bar{h}_l$.   In Eq.\ (\ref{Leff}),  we  have
included the bare Dirac bilinear $m \bar{N} N$.  As we will see below,
the effect  of this term  is to shift  the mass of the lowest-lying KK
state  by  an amount $m$.   In  principle,  we  could also  have added
another  Lorentz-   and  gauge-invariant fermionic    bilinear in Eq.\
(\ref{Leff}),  i.e.\ $M N^T C^{(5)-1}  N$, where $C^{(5)} = - \gamma_1
\gamma_3$.  The presence of the  latter operator is not very essential
for our phenomenological discussion.    In fact, this last term  drops
out, if one imposes invariance of  the Lagrangian (\ref{Leff}) under a
global transformation that  respects lepton number:  $N\to e^{i\theta}
N$, $L_l\to e^{-i\theta} L_l$ and $l_R\to e^{i\theta} l_R$.

We can  now express the 5-dimensional  two-component spinors $\xi$ and
$\eta$ of $N(x,y)$ in terms of a Fourier series expansion as follows:
\begin{eqnarray}
  \label{xi}
\xi (x,y) &=& \frac{1}{\sqrt{2\pi R}}\ \sum_{n=-\infty}^\infty\, \xi_n (x)\ 
                               \exp\bigg(\frac{iny}{R}\bigg)\,,\\
  \label{eta}
\eta (x,y) &=& \frac{1}{\sqrt{2\pi R}}\ \sum_{n=-\infty}^\infty\, \eta_n (x)\ 
                               \exp\bigg(\frac{iny}{R}\bigg)\,.
\end{eqnarray}
Substituting Eqs.\  (\ref{xi})  and  (\ref{eta}) into  the   effective
Lagrangian (\ref{Leff}) and then performing the $y$ integration yields
\begin{eqnarray}
  \label{LKK}
{\cal L}_{\rm eff} & = & {\cal L}_{\rm SM} (\Phi )\ +\ 
\sum_{n=-\infty}^\infty\, \bigg\{\, \bar{\xi}_n 
( i\bar{\sigma}^\mu \partial_\mu) \xi_n\ +\ 
\bar{\eta}_n ( i\bar{\sigma}^\mu \partial_\mu) \eta_n\ -\ 
\Big[\, \Big( m + \frac{in}{R}\, \Big)\, \xi_n \eta_{-n}\ +\
{\rm H.c.}\, \Big]\nonumber\\
&&
+\, \Big(\, \sum_{l=e,\mu,\tau}\,\bar{h}^{(n)}_l\, L_l\tilde{\Phi} \xi_n\ +\
{\rm H.c.}\,\Big)\, \bigg\}\, ,
\end{eqnarray}
where
\begin{equation}
  \label{hln}
\bar{h}^{(n)}_l\ =\  \frac{M_F}{M_{\rm P}}\ h_l\, 
\exp\bigg(\frac{ina}{R}\bigg)\, .
\end{equation}
As was first noticed  in \cite{ADDM,DDG}, the four-dimensional  Yukawa
couplings  $\bar{h}^{(n)}_l$  are naturally  suppressed by  the volume
factor $M_F/M_{\rm P}$ of the extra dimensions.

After spontaneous symmetry breaking (SSB), the effective Lagrangian of
the KK neutrino-mass matrix reads
\begin{eqnarray}
  \label{LKKmass}
{\cal L}^{\rm KK}_{\rm mass} &=& \Psi^T_+ {\cal M} \Psi_-\ +\ {\rm
H.c.},
\end{eqnarray}
where $\Psi^T_+ = (\nu_{lL}, \eta_0, \eta_1, \eta_{-1}, \dots, \eta_n,
\eta_{-n},  \dots)$,  $\Psi^T_-   = (\xi_0, \xi_{-1},    \xi_1, \dots,
\xi_{-n}, \xi_n, \dots)$ (with $n > 0$), and
\begin{equation}
  \label{MKK}
{\cal M}\ =\ \left( \begin{array}{ccccccc}
m^{(0)}_l & m^{(-1)}_l & m^{(1)}_l& \cdots & m^{(-n)}_l & m^{(n)}_l & \cdots \\
m & 0 & 0 & \cdots & 0 & 0 & \cdots \\
0 & m -\frac{i}{R} & 0 & \cdots & 0 & 0 & \cdots\\
0 & 0 & m +\frac{i}{R} & \cdots & 0 & 0 & \cdots\\
\vdots & \vdots & \vdots & \ddots & \vdots & \vdots & \cdots\\
0 & 0 & 0 &\cdots & m-\frac{in}{R} & 0 & \cdots \\
0 & 0 & 0 & \cdots & 0 & m+\frac{in}{R} & \cdots \\
\vdots & \vdots & \vdots & \vdots & \vdots & \vdots & \ddots 
\end{array} \right)\, ,
\end{equation}
where $m^{(n)}_l =   \bar{h}^{(n)}_l v/\sqrt{2}$, with  $l =  e,  \mu,
\tau$ and $\left< \Phi \right> = v/\sqrt{2}$.  Observe that the matrix
${\cal M}$  contains three rows   more than   the  rectangular case.   
However, it can be shown that  the additional rows correspond to three
massless Weyl spinors,  collectively  denoted as  $\nu_l$, and can  be
treated  independently of  the  rectangular part  of the neutrino-mass
matrix.  The   massless    Weyl  spinors  $\nu_l$ are    predominantly
left-handed and hence describe the observable neutrinos.

To make this very last point explicit, we will  first go from the weak
basis $\Psi_+$  to another rotated basis,   e.g.\ $\Psi^R_+$, which is
defined by the   unitary transformation:  $\Psi_+  = U^\nu  \Psi^R_+$,
where $(\Psi^R_+)^T =  (\nu_l, \eta^R_0, \eta^R_1, \eta^R_{-1}, \dots,
\eta^R_n, \eta^R_{-n}, \dots)$ and
\begin{equation}
  \label{Unu}
U^\nu\ =\ \left( \begin{array}{cc}
({\bf 1}_3 + \Xi^*\Xi^T)^{-1/2}  & \Xi^* ({\bf 1} + \Xi^T\Xi^*)^{-1/2} \\
-\Xi^T ({\bf 1}_3 + \Xi^*\Xi^T)^{-1/2} & ({\bf 1} + \Xi^T\Xi^*)^{-1/2} 
\end{array} \right)\, ,
\end{equation}
with 
\begin{equation}
  \label{Xi}
\Xi\ =\ \Big(\, \frac{m^{(0)}_l}{m}\ ,\ 
\frac{m^{(-1)}_l}{m -\frac{i}{R}}\ ,\ 
\frac{m^{(1)}_l}{m + \frac{i}{R}}\ , \
\cdots,\ 
\frac{m^{(-n)}_l}{m -\frac{in}{R}}\ ,\ 
\frac{m^{(n)}_l}{m + \frac{in}{R}}\ , \
\cdots\, \Big)\, ,
\end{equation}
and   $\Xi^*   ({\bf   1}   +  \Xi^T\Xi^*)^{-1/2}    =  ({\bf  1}_3  +
\Xi^*\Xi^T)^{-1/2}   \Xi^*$.  In   Eq.\ (\ref{Unu}),   the  root of  a
Hermitian  matrix, e.g.\ $H  = ({\bf 1}  + \Xi^T\Xi^*)$, is defined as
$H^{1/2} = U_H \hat{H}^{1/2}  U^\dagger_H$, where $U_H$ is the unitary
matrix that diagonalizes $H$, i.e.\ $\hat{H}  = U_H^\dagger H U_H$. It
is then easy to verify that $H^{1/2} H^{1/2} = H$, as it should be. In
the newly introduced weak  basis, the three upper  rows of the rotated
neutrino-mass  matrix  ${\cal   M}^R  =  (U^\nu)^T   {\cal  M}$ vanish
identically,   giving rise to  three massless   chiral fields $\nu_l$,
while the remainder of the  matrix assumes the usual rectangular  form
that describes  massive Dirac fields.  In  fact, for the case at hand,
one has $m^{(n)}_l \ll m$  and the massless  chiral fields $\nu_l$ are
predominantly left-handed, i.e.\ 
\begin{equation}
  \label{nul}
\nu_{lL}\ =\ ({\bf 1}_3 + \Xi^*\Xi^T)^{-1/2}\,
\Big(\,\nu_l\ +\ \Xi^* \Psi'^R_+\,\Big)\,,
\end{equation}
with $\nu^T_{l(L)} =  (\nu_{e(L)}, \nu_{\mu (L)}, \nu_{\tau (L)})$ and
$(\Psi'^R_+)^T =  (\eta^R_0, \eta^R_1,  \eta^R_{-1}, \dots,  \eta^R_n,
\eta^R_{-n}, \dots)$. In the limit $m\to  0$ discussed in \cite{ADDM},
there is a    level-crossing  effect and  $\eta_0$   becomes massless,
whereas one linear combination of the three  $\nu_l$ fields acquires a
small   Dirac   mass  of order $m^{(0)}_l$;     the  other  two linear
combinations orthogonal  to  the    last one remain     massless.  The
rectangular part of the matrix ${\cal  M}$, ${\cal M}_\chi$, which are
spanned  by  the  field   vectors  $\Psi'^R_+$ and   $\Psi_-$, can  be
diagonalized independently  by  a   bi-unitary transformation:  $V^T_+
{\cal M}_\chi V_- = \widehat{\cal M}_\chi$. We denote the resulting KK
mass eigenfields by $\chi^{(n)}$.   To leading order in $m^{(n)}_l/m$,
the mass eigenvalues of ${\cal M}_\chi$ are given by
\begin{equation}
  \label{MNhat} 
\widehat{\cal  M}_\chi\  \approx\ {\rm diag}\, \bigg(\,  m,\ 
\sqrt{m^2+\frac{1}{R^2}}\ ,\  \sqrt{m^2+\frac{1}{R^2}}\  ,\ \cdots,\ 
\sqrt{m^2+\frac{n^2}{R^2}}\  ,\  \sqrt{m^2+\frac{n^2}{R^2}}\ ,\ \cdots\,
\bigg)\, .
\end{equation}
In  this  approximation, up  to  phase factors,  the unitary  matrices
$V_\pm$ stay close to unity \cite{DS}.  Note that unlike $\chi^{(0)}$,
all  other massive Dirac   fields  fall into  degenerate pairs,  i.e.\ 
$m_{(n)} = m_{(-n)}$, with $n > 0$.  As we have mentioned above, since
the chiral neutrino $\nu_l$ is massless, the next-to-lightest state of
the neutrino mass spectrum, $\chi^{(0)}$, exhibits a mass gap of order
$m$, i.e.\ $m_{(0)} \approx m$.

In the following, we shall give  the Lagrangians \cite{SV} that govern
the  interactions of the neutrinos,  $\nu_l$ and $\chi^{(n)}$, and the
charged leptons, $l$, with the gauge  bosons, $W^\pm$ and $Z$, as well
as with their respective would-be Goldstone bosons, $G^\pm$ and $G^0$.
The interaction Lagrangians can be summarized as follows:
\begin{eqnarray}
  \label{Wint}
{\cal L}^{W^\pm}_{\rm int} \!\!&=&\!\! -\, \frac{g_w}{\sqrt{2}}\, W^{-\mu}
\sum_{l=e,\mu,\tau} \bigg(\, B_{l\nu_l}\, \bar{l}\gamma_\mu P_L \nu_l\
+\ \sum_{n=-\infty}^\infty B_{l,n}\, \bar{l}\gamma_\mu P_L
\chi^{(n)}\ +\ {\rm H.c.}\, \bigg)\, ,\\
  \label{Zint}
{\cal L}^Z_{\rm int} \!\!&=&\!\! -\, \frac{g_w}{2 c_w}\, Z^{\mu}\, \bigg[\,
\sum_{l,l'=e,\mu,\tau}
C_{\nu_l\nu_{l'}}\,\bar{\nu_l} \gamma_\mu P_L \nu_{l'}\ +\
\bigg( \sum_{l = e,\mu,\tau}\
\sum_{n=-\infty}^\infty C_{\nu_l,n}\, \bar{\nu_l} \gamma_\mu P_L
\chi^{(n)}\ +\ {\rm H.c.}\, \bigg)\nonumber\\
&& +\, 
\sum_{n,k=-\infty}^\infty C_{n,k}\, \bar{\chi}^{(n)} \gamma_\mu P_L
\chi^{(k)}\, \bigg]\, ,\\
  \label{Gplus}
{\cal L}^{G^\pm}_{\rm int} \!\!&=&\!\! -\, \frac{g_w}{\sqrt{2}M_W}\, G^-
\sum_{l=e,\mu,\tau} \bigg[\, B_{l\nu_l}\, m_l \bar{l}  P_L \nu_l\
+\ \sum_{n=-\infty}^\infty B_{l,n}\, \bar{l} \Big( m_lP_L - m_{(n)}P_R\Big)
\chi^{(n)}\nonumber\\
&&+\, {\rm H.c.}\, \bigg]\, ,\\
  \label{G0int}
{\cal L}^{G^0}_{\rm int} \!\!&=&\!\! \frac{ig_w}{2M_W}\, G^0\,
\bigg[\, \sum_{l=e,\mu,\tau} \bigg(\, 
\sum_{n=-\infty}^\infty C_{\nu_l,n}\, m_{(n)} \bar{\nu_l}P_R\chi^{(n)}\ 
+\ {\rm H.c.}\, \bigg)\nonumber\\
&& -\, \sum_{n,k=-\infty}^\infty C_{n,k}\, \bar{\chi}^{(n)} 
\Big( m_{(n)} P_L - m_{(k)} P_R\Big)\chi^{(k)}\, \bigg]\, ,
\end{eqnarray}
where $g_w$ is  the SU(2)$_L$ coupling constant, $c^2_w  = 1 - s^2_w =
M^2_W/M^2_Z$, $P_{L(R)} =  [1 -  (+)  \gamma_5]/2$ are  the  chirality
projection  operators, and $m_l$  and $m_{(n)}$ indicate the masses of
the charged leptons and  KK neutrinos, respectively. The matrices  $B$
and $C$ appearing in Eqs.\ (\ref{Wint})--(\ref{G0int}) are defined as
\begin{eqnarray}
  \label{Bmatr}
B_{l\nu_{k}} \!\!&=&\!\! \sum_{i=e,\mu,\tau} V^l_{li} U^\nu_{ik}\quad
({\rm with}\ k = e,\mu,\tau), \qquad
B_{l,n}\ =\ \sum_{i=e,\mu,\tau} V^l_{li} U^\nu_{i,n}\, ,\\
  \label{Cmatr}
C_{\nu_l\nu_{l'}} \!\!&=&\!\! 
\sum_{k= e,\mu,\tau} B_{k\nu_l}B^*_{k\nu_{l'}}\,,\quad 
C_{\nu_l ,n}\ =\ \sum_{k= e,\mu,\tau} B_{k\nu_l}B^*_{k,n}\, ,\quad
C_{n,m}\ =\ \sum_{k= e,\mu,\tau} B_{k,n}B^*_{k,m}\, ,\quad
\end{eqnarray}
where $V^l$ is a unitary matrix  that occurs in the diagonalization of
the   charged lepton  mass  matrix.   {}From  Eqs.\ (\ref{Bmatr})  and
(\ref{Cmatr}),   it  is   interesting  to  observe    that  the  gauge
interactions  between the SU(2)$_L$ doublets $L_l$  and  the KK states
$\chi^{(n)}$ are suppressed by a factor $m^{(n)}_l/m_{(n)}$, while the
corresponding couplings to   two KK neutrino states   $\chi^{(n)}$ and
$\chi^{(m)}$ are further suppressed by the ratio $m^{(n)}_lm^{(m)}_l /
(m_{(n)}m_{(m)})$.

\setcounter{equation}{0}
\section{Tree-level constraints}

The interaction Lagrangians  (\ref{Wint})--(\ref{G0int})  give rise to
important phenomena of new physics both at  the tree level and beyond.
In this section, we will determine the new-physics contributions of KK
neutrinos to  electroweak observables, which  occur at the tree level,
and  so derive  limits on the  fundamental Planck  scale $M_F$ and the
mixing parameters of the theory.

The most striking feature  of the  higher-dimensional scenario is  the
loss of lepton universality  in  electroweak processes involving  $W$-
and $Z$-boson interactions.  However, KK  neutrinos  may also lead  to
observable modifications of the  muon lifetime, the invisible width of
the  $Z$  boson,  the  cross section of   the  $\nu  e$ deep inelastic
scatterings, etc.   For our analysis, it  proves useful  to define the
mixing parameters, which were first introduced by Langacker and London
\cite{LL},
\begin{eqnarray}
  \label{snul}
(s^{\nu_l}_L)^2 &\equiv& 1\, -\, 
\sum_{l' = e,\mu,\tau}\, |B_{l\nu_{l'}}|^2\ =\ 
\sum_{n=-\infty}^\infty  |B_{l,n}|^2\nonumber\\
&=& [V^l ({\bf 1}_3 + \Xi^*\Xi^T )^{-1/2}
(\Xi^*\Xi^T) ({\bf 1}_3 + \Xi^*\Xi^T )^{-1/2} V^{l\dagger}]_{ll}\, .
\end{eqnarray}
In   order  to  evaluate  the last    equality on  the  RHS  of   Eq.\ 
(\ref{snul}), we  will  approximate the sum  over the  KK states by  a
higher-dimensional  energy integral, which  has an  upper ultra-violet
(UV)  cutoff  at   $M_F$,  above which  string-threshold  effects  are
expected to become   more  important.  To be   precise,  we make   the
replacement
\begin{equation}
  \label{sumKK}
\sum_{n=-\infty}^\infty\ \to \ S_\delta R^\delta \int^{M_F}_0
E^{\delta -1} dE\, ,
\end{equation}
where $S_\delta  = 2\pi^{\delta/2}  /\Gamma(\frac{\delta}{2})$ is  the
surface  area of  a    $\delta$-dimensional  sphere of   unit  radius.
Furthermore,  we consider that    the charged  lepton mass  matrix  is
diagonal and non-negative,  i.e.\ $V^l =  {\bf 1}_3$. Then, to leading
order  in   $(h_l  v)/M_F$,  we  find  that   $(s^{\nu_l}_L)^2 \approx
(\Xi^*\Xi^T)_{ll}$, with
\begin{equation}
  \label{XiXi}
(\Xi^*\Xi^T)_{ll}\ =\ |m^{(0)}_l|^2\, S_\delta R^\delta\, \int^{M_F}_0 
\frac{ dE\, E^{\delta -1}}{m^2 + E^2}\, =\,
\left\{ \begin{array}{l} \frac{\displaystyle 
\pi h^2_l v^2}{\displaystyle M^2_F}\, 
\ln\bigg(\displaystyle \frac{M^2_F + m^2}{\displaystyle m^2}\bigg)\, ,
\quad {\rm for}\ \delta=2 \\ \\
\frac{\displaystyle S_\delta}{\displaystyle \delta - 2}\,
\frac{\displaystyle h^2_l v^2}{\displaystyle M^2_F}\, \Big[\,1 +\, 
{\cal O}(m^2/M^2_F)\Big],\quad 
{\rm for}\ \delta > 2\, .
\end{array}\ \right.
\end{equation}
Equation   (\ref{XiXi}) shows that  deviations   of  the $Wl\nu$   and
$Z\nu\nu$ couplings from their  SM values are logarithmically enhanced
in a theory with two compact dimensions \cite{FP}.  In the limit $m\to
0$, in which the scenario of Ref.\ \cite{ADDM}  is recovered, one must
simply  replace  the  logarithm  $\ln[(M^2_F   +  m^2)/m^2]$  by  $\ln
(M^2_{\rm P}/M^2_F)$ in Eq.\  (\ref{XiXi}).  Furthermore, we find that
the parameters $(s^{\nu_l}_L)^2$  are  not  suppressed by the   volume
factor $M^2_F/M^2_P$, as would have been the case if we had not summed
over all  the KK states \cite{FP}.  In  fact, after summation over the
KK states,  we   obtain  an  effective  theory  in  which  the  Yukawa
interactions are mediated by couplings  $h_l$ of  order unity.  As  we
will see in   Sections 4 and 5,   the latter gives  rise to observable
non-decoupling effects at the one-loop electroweak order, and can lead
to  more severe limits   on the parameters  of  the theory  than those
considered here.

The mixing parameters $(s^{\nu_l}_L)^2$ may   now be constrained by  a
number  of experimental results,  which  are  obtained from: (i)   the
precise measurement of the muon width $\Gamma(\mu \to e\nu \nu)$; (ii)
the  neutrino counting   at   the  $Z$  peak;  (iii)   charged-current
universality in pion decays, i.e.\ $\Gamma(\pi \to e\nu)/\Gamma(\pi\to
\mu\nu)$;  (iv)  charged-current  universality   in tau  decays, i.e.\ 
$B(\tau \to e \nu \nu)/B(\tau \to \mu  \nu\nu)$.  In the following, we
shall discuss in  more detail the constraints  obtained from limits on
the new-physics phenomena mentioned above.

\noindent
(i) {\em Precise measurement of the muon lifetime}. The presence of KK
neutrinos leads to a modification of the muon width $\Gamma (\mu \to e
\nu \nu)$, when compared to the SM result  $\Gamma_{\rm SM} (\mu \to e
\nu \nu)$. In particular, we have
\begin{equation}
  \label{Gmu}
1\, -\ \frac{\Gamma (\mu \to e\nu\nu)}{\Gamma_{\rm SM} (\mu \to e\nu\nu)}\ 
=\ (s^{\nu_e}_L)^2\, +\, (s^{\nu_\mu}_L)^2\  \stackrel{<}{{}_\sim}\ 0.01\, . 
\end{equation}
The upper limit derived above is  very conservative, in the sense that
we also  estimated  the impact of  neglecting  high-order  terms.  The
major new-physics contributions  come from one-loop corrections to the
$W$-boson propagator.   These  one-loop corrections  are quantified by
the electroweak    oblique parameters,  such   as  $S$,  $T$  and  $U$
\cite{STU}, and will be discussed in Section 5.  Furthermore, one-loop
vertex effects generically introduce further corrections to the mixing
parameters   $(s^{\nu_l}_L)^2$,   which  could    be   of  order  15\%
\cite{BAKAP}.

\noindent
(ii)    {\em  Neutrino  counting     at  the   $Z$  peak}.     In  the
higher-dimensional model under study, the coupling of the $Z$ boson to
the massless   neutrinos   $\nu_l$ is  reduced   by   a  factor  $1  -
(s^{\nu_l}_L)^2$ relative to the SM case.  The  latter would result in
an observable change of the $Z$-boson invisible width at LEP, which is
translated into the upper limit \cite{PDG}
\begin{equation}
  \label{Zinv}
1\, -\, \frac{\Gamma ( Z \to \nu\nu)}{ \Gamma_{\rm SM}( Z \to \nu\nu)}\
=\ (s^{\nu_e}_L)^2\, +\, (s^{\nu_\mu}_L)^2\, +\, (s^{\nu_\tau}_L)^2\
<\ 0.034\, ,
\end{equation}
at the $2\sigma$   level. Equations (\ref{Gmu}) and  (\ref{Zinv}) give
rather  reliable    upper   limits    on    the  absolute    size   of
$(s^{\nu_l}_L)^2$.   The    remaining   observables  measure  possible
deviations from lepton universality.

\noindent
(iii) {\em Charged-current universality  in pion decays}. One  may now
define an  analogous   observable    in order  to     probe  $e$-$\mu$
universality, i.e.
\begin{equation}
  \label{Rpiviol}
1\, -\, \frac{R_\pi}{R^{\rm SM}_\pi} \ =\ (s^{\nu_e}_L)^2\, 
-\, (s^{\nu_\mu}_L)^2\, ,
\end{equation}
where
\begin{equation}
  \label{Rpi}
R_\pi\ =\ \frac{\Gamma ( \pi \to e\nu)}{ \Gamma ( \pi \to \mu\nu) }
\end{equation}
and $R^{\rm  SM}_\pi$ is the SM  result.  The experimental analysis of
the observable in   Eq.\  (\ref{Rpiviol}) yields the  $2\sigma$  upper
limit  \cite{DAB}\footnote{Most recently,  it  was noticed   \cite{DK}
that, if $m=0$ in Eq.\ (\ref{Leff}), the  lower bound on $M_F$ that is
deduced from charged-current universality  in pion decays can  be very
tight, when phase-space effects of KK states lighter than the pion are
taken into account. There are two ways to  evade such a limit: (i) one
may consider  $m\sim M_W \gg m_\pi$,  so the next-to-lightest KK state
is  much  heavier than  $\pi^+$;  (ii)   one could   assume a kind  of
$\mu$-$e$ universality, with $m=0$.}
\begin{equation}
  \label{Rpilimit}
(s^{\nu_e}_L)^2\, -\, (s^{\nu_\mu}_L)^2\ <\ 0.003 \pm 0.006\, .
\end{equation}

\noindent
(iv)     {\em Charged-current   universality in     tau decays}.  Yet,
universality-breaking    effects   in  the   leptonic  sector  through
charged-current interactions can be  examined in the tau decays, $\tau
\to e\nu\nu$ and  $\tau  \to \mu\nu\nu$. For  this purpose,   we first
define the quantities:
\begin{equation}
  \label{Rtemu}
R_{\tau e}\ =\ \frac{ \Gamma ( \tau \to e \nu \nu )}{\Gamma (\mu \to e
\nu \nu )}\, ,\qquad
R_{\tau \mu}\ =\ 
\frac{ \Gamma ( \tau \to \mu \nu \nu )}{\Gamma (\mu \to e
\nu \nu )}\, .
\end{equation}
By analogy, the non-SM contributions  to the above two observables are
constrained by \cite{DAB}
\begin{eqnarray}
  \label{Rteexp}
1\, -\, \frac{R_{\tau e}}{R^{\rm SM}_{\tau e}} &=& 
(s^{\nu_\tau}_L)^2\, -\, (s^{\nu_e}_L)^2\ <\ 0.032 \pm 0.048\, ,\\
  \label{Rmueexp}
1\, -\, \frac{R_{\tau \mu}}{R^{\rm SM}_{\tau \mu}} & =& 
(s^{\nu_\tau}_L)^2\, -\, (s^{\nu_\mu}_L)^2\ <\ 0.040 \pm 0.048\, ,
\end{eqnarray}
at the $2\sigma$     level.   More recent  experimental   analyses  of
$\tau$-$e$  and  $\tau$-$\mu$  universality  \cite{APich}   lead to  a
significant improvement of the above $2\sigma$ upper limits, i.e.\
\begin{equation}
  \label{Rtemunew}
(s^{\nu_\tau}_L)^2\, -\, (s^{\nu_e}_L)^2\quad {\rm or} \quad
(s^{\nu_\tau}_L)^2\, -\, (s^{\nu_\mu}_L)^2\ <\ 0.012\, .
\end{equation}
These    last   limits  are competitive    with   those  obtained from
considerations of charged-current universality in pion decays.

%%%%%%%%%%%%%%%%%%%%%%%%%%%%%%%%%%%%%%%%%%%%%%%%%%%%%%%%%%%%%%%%%%%%%%%%%%
%   TABLE 1
%%%%%%%%%%%%%%%%%%%%%%%%%%%%%%%%%%%%%%%%%%%%%%%%%%%%%%%%%%%%%%%%%%%%%%%%%%
\begin{table}[t]

\begin{center}

\begin{tabular}{|c||cc||cc|}
\hline
 & $h_e\ =\ h_\mu$ &\hspace{-1.7cm}$=\ h_\tau\ =\ h$ & 
\hspace{1.cm}$h_\mu =0$ &\hspace{-1.3cm}and $h_e = h_\tau$ \\
Observable & Lower limit & 
\hspace{-0.6cm} on $M_F/h\ [\,{\rm TeV}\,]$ & Lower limit 
&\hspace{-0.6cm} on $M_F/h_\tau\ [\,{\rm TeV}\,]$ \\ 
& $\delta =2$ & $\delta > 2$ & $\delta =2 $ & $\delta > 2$ \\
\hline\hline
&&&&\\
$1 - \frac{\displaystyle \Gamma(\mu\to e\nu\nu )}{\displaystyle 
\Gamma_{\rm SM} (\mu\to e\nu\nu )}$
 & $8.9\,\ln^{1/2}\frac{\displaystyle M_F}{\displaystyle m}$ &
$\frac{\displaystyle 3.5\, S^{1/2}_\delta}{\displaystyle \sqrt{\delta -2} }$& 
$6.3\,\ln^{1/2}\frac{\displaystyle M_F}{\displaystyle m}$& 
$\frac{\displaystyle 2.5\, S^{1/2}_\delta}{\displaystyle \sqrt{\delta -2} }$\\
&&&&\\
\hline
&&&&\\
$1 - \frac{\displaystyle \Gamma(Z\to \nu\nu )}{\displaystyle 
\Gamma_{\rm SM} ( Z\to \nu\nu )}$
 & $5.9\,\ln^{1/2}\frac{\displaystyle M_F}{\displaystyle m}$ &
$\frac{\displaystyle 2.4\, S^{1/2}_\delta}{\displaystyle \sqrt{\delta -2}}$& 
$4.8\,\ln^{1/2}\frac{\displaystyle M_F}{\displaystyle m}$& 
$\frac{\displaystyle 1.9\, S^{1/2}_\delta}{\displaystyle \sqrt{\delta -2}}$\\
&&&&\\
\hline
&&&&\\
$1 - \frac{\displaystyle R_\pi }{\displaystyle R^{\rm SM}_\pi }$
        & $-$ & $-$ & 
$18.7\,\ln^{1/2}\frac{\displaystyle M_F}{\displaystyle m}$& 
$\frac{\displaystyle 7.5\, S^{1/2}_\delta}{\displaystyle \sqrt{\delta -2}}$\\ 
&&&&\\
\hline
&&&&\\
$1 - \frac{\displaystyle R_{\tau \mu}}{\displaystyle R^{\rm SM}_{\tau \mu} }$
 & $-$ & $-$ & $5.7\,\ln^{1/2}\frac{\displaystyle M_F}{\displaystyle m}$& 
$\frac{\displaystyle 2.3\, S^{1/2}_\delta}{\displaystyle \sqrt{\delta -2}}$\\
&&&&\\
\hline
\end{tabular}
\end{center}

\caption{Limits  on $M_F$ and $h_l$ at  the $2\sigma$ level.  The case
$m=0$   is  obtained by  replacing $\ln    (M_F/m)$  with $\ln (M_{\rm
P}/M_F)$.}\label{Tab1}

\end{table}

In Table \ref{Tab1} we show the lower limits  on the fundamental scale
$M_F$, as well as  the upper limits  on the  higher-dimensional Yukawa
couplings $h_l$, for    two  representative scenarios. In   the  first
scenario,  we have considered  the complete universality of the Yukawa
couplings, i.e.\ $h_e = h_\mu = h_\tau =  h$.  In the second scenario,
we assume that the $\mu$ lepton does not mix with the singlet neutrino
$N(x,y)$, and $h_e = h_\tau$. Of  course, we might have considered the
converse case, in which electron does not mix with $N(x,y)$ instead of
the muon, but  the predictions  that we would  obtain  then would  not
differ much. In both scenarios, we assume  that the singlet Dirac mass
$m$ is of order $M_W$.  {}From Table \ref{Tab1}, we see that precision
measurements  on muon  lifetime    offer the  most  sensitive test  of
lepton-flavour  universality in the  first scenario.  For instance, if
$h =1$ and $m=100$ GeV ($m = 0$), we derive the limits:
\begin{equation}
  \label{MFI}
M_F\ \stackrel{>}{{}_\sim}\ 20\ (65)\,,\ 12.5\,,\ 9.9\ {\rm TeV}\,, 
\end{equation}
for $\delta = 2$, 3  and 6 large  compact dimensions, respectively. In
the second  scenario, the best    bounds are obtained by  looking   at
deviations from charged-current universality  in pion decays. Thus, if
we take $h_e = h_\tau = 1$ and $m = 100$  GeV ($m = 0$), the following
lower bounds on $M_F$ may be derived:
\begin{equation}
  \label{MFII}
M_F\ \stackrel{>}{{}_\sim}\ 45\ (102)\,,\ 26\,,\ 21\ {\rm TeV}\,, 
\end{equation}
for $\delta = 2$, 3 and 6 large compact dimensions, respectively.  The
very same limits apply to a third possible scenario, with $h_e = h_\mu
=  0$ and $h_\tau =  1$. For comparison, we  note  that the respective
lower limits on   $M_F$ derived from  upper  bounds  on the  invisible
$Z$-boson width  (cf.\ Eq.\  (\ref{Zinv}))  in the  third scenario are
weaker, i.e.
\begin{equation}
  \label{MFIII}
M_F\ \stackrel{>}{{}_\sim}\ 8.2\ (23)\,,\ 4.7\,,\ 3.8\ {\rm TeV}\, ,
\end{equation}
for 2, 3 and 6 large compact dimensions.  Notice that the lower limits
on  $M_F$ displayed in  Eqs.\ (\ref{MFI})--(\ref{MFIII}) for the three
different scenarios increase linearly with the Yukawa couplings $h_l$.
There may be additional tree-level processes  that could constrain the
$e$-$\mu$-$\tau$ sectors,  e.g.\  observable change   of the $\nu   e$
deep-inelastic-scattering  data.  However, the additional  constraints
turn out to be comparable to those we  listed above.  In Section 5, we
shall see  that one-loop non-decoupling   effects of KK  neutrinos can
lead to   much    stronger   bounds  than   those    given   by  Eqs.\
(\ref{MFI})--(\ref{MFIII}).

\setcounter{equation}{0}
\section{Cumulative non-decoupling effect of Kaluza--Klein neutrinos}

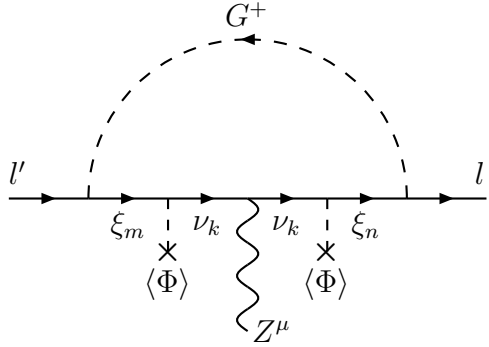
\begin{figure}

\begin{center}
\begin{picture}(200,150)(0,0)
\SetWidth{0.8}

\ArrowLine(0,50)(30,50)\ArrowLine(30,50)(60,50)
\ArrowLine(60,50)(90,50)\ArrowLine(90,50)(120,50)
\ArrowLine(120,50)(150,50)\ArrowLine(150,50)(180,50)
\DashArrowArc(90,50)(60,0,180){5}\Photon(90,50)(90,0){4}{3}
\DashLine(60,50)(60,30){4}\DashLine(120,50)(120,30){4}

\Text(90,120)[]{$G^+$}\Text(92,0)[l]{$Z^\mu$}\Text(0,60)[l]{$l'$}
\Text(180,60)[r]{$l$}\Text(45,40)[]{$\xi_m$}\Text(135,40)[]{$\xi_n$}
\Text(75,40)[]{$\nu_k$}\Text(105,40)[]{$\nu_k$}
\Text(60,18)[]{$\left< \Phi \right>$}\Text(60,30)[]{\boldmath $\times$}
\Text(120,18)[]{$\left< \Phi \right>$}\Text(120,30)[]{\boldmath $\times$}

\end{picture}\\
\end{center}

\caption{Feynman graph related to the dominant non-decoupling
part of the effective $Zll'$ coupling in the Feynman gauge.}\label{fig:1}

\end{figure}

It  is now  very    instructive    to explicitly    demonstrate    how
higher-dimensional Yukawa  couplings of  order  unity give rise   to a
non-decoupling   effect  mediated   by  KK  neutrinos in   electroweak
processes.       As   an     example,   we       will  consider    the
lepton-flavour-violating vertex  $Zll'$, shown in  Fig.\ \ref{fig:1}.  
Similar  non-decoupling  effects occur in    box diagrams involving KK
neutrinos.  A more quantitative discussion  of one-loop constraints on
the parameters of the theory follows in the next section.

%******************************************************************
%   Figure 1
%******************************************************************

Adopting  the Feynman gauge for  simplicity, the dominant contribution
to the lepton-flavour-violating  vertex $Zll'$ comes from  the Feynman
diagram displayed in  Fig.\ \ref{fig:1}.   The non-decoupling part  of
the effective $Zll'$ coupling may then be given by
\begin{equation}
  \label{TZll}
{\cal T}(Zll')\ =\ \sum_{n,m = -\infty}^\infty 
                                       {\cal T}_{(n,m)} (Zll')\, ,
\end{equation}
where 
\begin{eqnarray}
  \label{TZllnm}
{\cal T}_{(n,m)} (Zll') \!\!&=&\!\! \frac{g_w}{4c_w}\, 
\bar{h}^{(n)}_l\, \bar{h}^{(m)*}_{l'}\, \bigg( \sum\limits_{k=e,\mu,\tau}
\bar{h}^{(n)*}_k \bar{h}^{(m)}_k\bigg)\, v^2\nonumber\\
&&\times\, \int\! 
\frac{d^4k}{(2\pi )^4 i}\, P_R \frac{1}{\not\! k - m_{(n)}}\, P_L\,
\frac{1}{\not\! k}\, \gamma_\mu P_L\, \frac{1}{\not\! k}\, P_R\,
\frac{1}{\not\! k - m_{(m)}}\, P_L\, \frac{1}{k^2 - M^2_W}\\
&&\hspace{-2.5cm}=\ (\gamma_\mu P_L)\,\frac{g_w}{64\pi^2 c_w}\,
\frac{M^4_F}{M^4_{\rm P}}\, \frac{h_l h_{l'} v^2}{M^2_W}\,
\bigg( \sum\limits_{k=e,\mu,\tau} h^2_k \bigg)\,
\frac{\lambda_n \lambda_m \ln (\lambda_n/\lambda_m)\, +\,
\lambda_m \ln\lambda_m\, -\, \lambda_n \ln\lambda_n }{(\lambda_m -\lambda_n)
(1 - \lambda_m ) (1 - \lambda_n)}\ ,\nonumber
\end{eqnarray}
with $\lambda_n = m^2_{(n)}/M^2_W$.  {}From the  last equality in Eq.\
(\ref{TZllnm}), one naively finds that the individual KK contributions
${\cal T}_{(n,m)} (Zll')$ are indeed tiny, since they are proportional
to  the   volume-dependent   suppression  factor $M^4_F/M^4_{\rm    P}
\approx 10^{-64}$, for $M_F \approx 1$ TeV. 

Let us now evaluate the  sum over the KK states  in Eq.\ (\ref{TZll}). 
The loop function in the last equality of Eq.\ (\ref{TZllnm}) receives
its biggest  support from KK  neutrinos  heavier than the  $W$ boson.  
Thus,    converting  the  double sum   over    the KK states  in  Eq.\ 
(\ref{TZll}) into a double integral (cf.\ Eq.\ (\ref{sumKK})) that has
an infra-red (IR) cutoff at $M_W \sim m$, we find
\begin{eqnarray}
  \label{TZll1}
{\cal T} (Zll') &\approx& -\, (\gamma_\mu P_L)\,
\frac{g_w}{64\pi^2 c_w}\, \frac{M^4_F}{M^4_{\rm P}}\, 
\frac{h_l h_{l'} v^2}{M^2_W}\, 
\bigg(\sum\limits_{k=e,\mu,\tau} h^2_k \bigg)\nonumber\\
&&\times\, \frac{M^2_WS^2_\delta}{4M^2_F}\, 
(M_F R)^{2\delta}\!\!
\int\limits_{m^2/M^2_F}^1\!\!\! dz\, z^{\frac{\delta}{2} - 1}\, \!\!
\int\limits_{m^2/M^2_F}^1\!\!\! dw\, w^{\frac{\delta}{2} - 1}\
\frac{\ln (z/w)}{z-w}\ .
\qquad
\end{eqnarray}
Employing  the  relation   (\ref{Gauss}) that  governs the  parameters
$M_F$, $M_{\rm  P}$ and  $R$,  we  observe  that the huge  suppression
factor $M^4_F/M^4_{\rm P}$ multiplying the individual KK contributions
is cancelled  against the total number  of the active KK states, $(M_F
R)^{2\delta}$!  The  double integral  in Eq.\ (\ref{TZll1})   takes on
values in  the range  1--$10^{-2}$, for $2  \le  \delta \le  6$.  More
precisely, to leading order  in $\varepsilon\equiv m^2/M^2_F\ll 1$, we
have
\begin{equation}
  \label{Idelta}
I_\delta\ \equiv\ \frac{1}{4}\, 
\int\limits_{\varepsilon}^1\! dz\, z^{\frac{\delta}{2} - 1}\, \!\!
\int\limits_{\varepsilon}^1\! dw\, w^{\frac{\delta}{2} - 1}\
\frac{\ln (z/w)}{z-w}\ =\,
\left\{ \begin{array}{l} 
\frac{1}{12}\pi^2\,,\quad {\rm for}\ \delta=2\\ 
\frac{1}{8}\pi^2 - 1 \,,\quad {\rm for}\ \delta=3\\
\frac{1}{36}\pi^2 - \frac{1}{6} \,,\quad {\rm for}\ \delta=4\\
\frac{1}{16}\pi^2 - \frac{5}{9}\,,\quad {\rm for}\ \delta=5\\
\frac{1}{60}\pi^2 - \frac{1}{8} \,,\quad {\rm for}\ \delta=6 
\end{array}\ \right.
\end{equation}
Then, the effective $Zll'$ coupling may be determined by
\begin{eqnarray}
  \label{TZll2}
{\cal T} (Zll') &\approx & -\,(\gamma_\mu P_L)\,
\frac{g_w \alpha_w }{16\pi c_w}\, \frac{h_l h_{l'} v^2}{M^2_F}\, 
\bigg( \sum\limits_{k=e,\mu,\tau}\frac{ h^2_k v^2}{M^2_W}\bigg)\, 
S^2_\delta\, I_\delta\nonumber\\
&=& -\,(\gamma_\mu P_L)\, \frac{g_w \alpha_w}{16\pi c_w}\, 
s_L^{\nu_l} s_L^{\nu_{l'}}\, 
\bigg[ \sum_{k=e,\mu,\tau} (s_L^{\nu_k})^2\bigg]\,
\frac{M^2_F}{M^2_W}\ d_\delta\,,
\end{eqnarray}
where  $\alpha_w = g^2_w/(4\pi)$  is    the SU(2)$_L$ fine   structure
constant,  $(s_L^{\nu_l})^2$  are mixing    parameters that  have been
estimated in Eq.\ (\ref{XiXi}), and $d_\delta$ are dimension-dependent
correction factors that take the values: 
\begin{eqnarray}
  \label{ddel}
d_2 \!\!&=&\!\! \frac{\pi^2}{12\,\ln^2 (M_F/m)}\ \approx\ 
\frac{0.822}{\ln^2(M_F/m)}\ ,\qquad 
d_3\ =\  \frac{\pi^2}{8} - 1\ \approx\ 0.234\,,\nonumber\\
d_4 \!\!&=&\!\! \frac{\pi^2}{9} - \frac{2}{3}\ \approx\ 0.430 \,, 
\qquad d_5\ =\ \frac{9}{16}\pi^2 - 5\ 
\approx\ 0.552\,,\nonumber\\
d_6 \!\!&=&\!\! \frac{4\pi^2}{15} - 2\ \approx\ 0.632\, .
\end{eqnarray}
{}From Eq.\  (\ref{TZll2}),  we   observe that  the strength   of  the
effective $Zll'$ coupling increases with the {\em fourth} power of the
higher-dimensional Yukawa couplings, $h_l$, while it only decreases as
$1/M^2_F$. Thus, the  lower bounds on $M_F$,   which are derived  from
limits  on new-physics signals    mediated by $Z$-boson  interactions,
increase {\em quadratically} with $h_l$.

One  might   now  raise  the   question  whether  such   a  cumulative
non-decoupling effect  of KK neutrinos could  be dramatically weakened
if neutrinos would live only in a $\delta$-dimensional subspace of the
whole  $n$-dimensional  space  that gravitational  interactions  would
propagate, i.e.\ $\delta  < n$.  Nevertheless, it is  not difficult to
see  that this  is not  true.  In  this case,  the volume  factor that
suppresses the Yukawa  couplings $h_l$ is $1/(M_FR)^{\delta/2} \approx
(M_F/M_{\rm   P})^{\delta/n}$  \cite{ADDM}.    This  suppresion-volume
factor occuring in the $h^4_l$-dependent terms will now cancel exactly
against the number of the  KK-neutrino states in the loop, i.e.\ $(M_F
R)^{2\delta}  \approx  (M_{\rm  P}/M_F)^{4\delta/n}$.  Therefore,  the
analytic results  derived in  Eqs.\ (\ref{TZll2}) and  (\ref{ddel}) do
not  directly depend  on the  number $n$  of dimensions  of  the space
experienced  by  gravity.   The  case  $\delta =  1$,  which  was  not
considered,  is  not phenomenologically  viable,  since  for the  most
conservative  case  with  $\delta  =  1$  and $n  =  2$,  one  obtains
unacceptably large Yukawa couplings of order $10^{-7}$ for order-unity
higher-dimensional  Yukawa couplings  $h_l$, giving  rise  to neutrino
masses in the keV range.

An analogous   cumulative non-decoupling phenomenon,  with a kinematic
dependence related to the  effective  $Zll'$ coupling, takes place  in
box   diagrams,  which involve KK   modes   and two oppositely charged
would-be  Goldstone bosons, $G^\pm$, in  the  loop. Finally, we should
remark that the effective  $\gamma ll'$ vertex considered in \cite{FP}
only scales quadratically with the Yukawa  coupling $h_l$.  As we will
see in  the   next section, the   bounds  on $M_F$  derived  from  the
non-observation of photonic muon decays, $\mu \to e \gamma$, are found
to be less restrictive than those obtained  from $\mu \not\!  \to eee$
and the absence of $\mu$-to-$e$-conversion events in nuclei.

\setcounter{equation}{0}
\section{One-loop constraints}

After  having  gained  some insight  of  the cumulative non-decoupling
mechanism of  KK neutrinos, we shall  now present analytic results for
the      most     precisely   tested   lepton-flavour-violating    and
universality-breaking  processes  that  involve   the $W^\pm$  and  $Z$
bosons,  and the $e$,  $\mu$  and  $\tau$ leptons.  These  electroweak
processes, which are  strictly forbidden in the  SM, are induced by KK
states at the one-loop electroweak order.   Based on these results, we
are then able to derive very stringent limits on the parameters of the
theory.     Specifically,  we  confront     the  predictions   of  the
higher-dimensional model with experimental  data for the following set
of observables of new physics: (i) photonic decays of the muon and tau
leptons:    $\mu  \to e\gamma$,   $\tau\to    e\gamma$  and  $\tau \to
\mu\gamma$; (ii) decays  of the  $\mu$ and  $\tau$ leptons into  three
lighter charged   leptons, e.g.\  $\mu\to eee$, $\tau\to   eee$, etc.;
(iii)     coherent    $\mu$-to-$e$     conversion    in  nuclei;  (iv)
lepton-flavour-violating   decays   of  the   $Z$  boson, $Z\to  ll'$,
universality-breaking effects among different diagonal leptonic decays
of  the  $Z$ boson,   $Z\to ll$,  and   among its  associate  leptonic
asymmetries; (v) the electroweak oblique  parameters $S$, $T$ and  $U$
\cite{STU}, and especially $T \propto \delta\rho$ \cite{Velt}.

\subsection{Photonic decays of the $\mu$ and $\tau$ leptons}

\begin{figure}

\begin{center}
\begin{picture}(360,400)(0,0)
\SetWidth{0.8}

\ArrowLine(0,360)(20,360)\ArrowLine(60,360)(80,360)
\GCirc(40,360){20}{0.7}\Photon(40,340)(40,320){3}{2}
\Text(0,365)[b]{$l$}\Text(80,365)[b]{$l'$}
\Text(42,320)[l]{$Z,\gamma$}

\Text(100,360)[]{$=$}

\ArrowLine(120,360)(140,360)\ArrowLine(180,360)(200,360)
\ArrowLine(140,360)(180,360)\Text(160,367)[b]{$\chi^{(n)}$}
\DashArrowArc(160,360)(20,180,270){3}\PhotonArc(160,360)(20,270,360){2}{3}
\Text(145,340)[r]{$G^-$}\Text(175,340)[l]{$W^-$}
\Photon(160,340)(160,320){3}{2}
\Text(120,365)[b]{$l$}\Text(200,365)[b]{$l'$}
\Text(162,320)[l]{$Z,\gamma$}
\Text(160,300)[]{\bf (a)}

\ArrowLine(240,360)(260,360)\ArrowLine(300,360)(320,360)
\ArrowLine(260,360)(300,360)\Text(280,367)[b]{$\chi^{(n)}$}
\DashArrowArc(280,360)(20,270,360){3}\PhotonArc(280,360)(20,180,270){2}{3}
\Text(265,340)[r]{$W^-$}\Text(295,340)[l]{$G^-$}
\Photon(280,340)(280,320){3}{2}
\Text(240,365)[b]{$l$}\Text(320,365)[b]{$l'$}
\Text(282,320)[l]{$Z,\gamma$}
\Text(280,300)[]{\bf (b)}

\ArrowLine(0,260)(20,260)\ArrowLine(60,260)(80,260)
\ArrowLine(20,260)(60,260)\Text(40,267)[b]{$\chi^{(n)}$}
\PhotonArc(40,260)(20,180,270){2}{3}\PhotonArc(40,260)(20,270,360){2}{3}
\Text(25,240)[r]{$W^-$}\Text(55,240)[l]{$W^-$}
\Photon(40,240)(40,220){3}{2}
\Text(0,265)[b]{$l$}\Text(80,265)[b]{$l'$}
\Text(42,220)[l]{$Z,\gamma$}
\Text(40,200)[]{\bf (c)}

\ArrowLine(120,260)(140,260)\ArrowLine(180,260)(200,260)
\ArrowLine(140,260)(180,260)\Text(160,267)[b]{$\chi^{(n)}$}
\DashArrowArc(160,260)(20,180,270){3}\DashArrowArc(160,260)(20,270,360){2}
\Text(145,240)[r]{$G^-$}\Text(175,240)[l]{$G^-$}
\Photon(160,240)(160,220){3}{2}
\Text(120,265)[b]{$l$}\Text(200,265)[b]{$l'$}
\Text(162,220)[l]{$Z,\gamma$}
\Text(160,200)[]{\bf (d)}

\ArrowLine(240,260)(260,260)\ArrowLine(300,260)(320,260)
\Photon(260,260)(300,260){2}{4}\Text(280,267)[b]{$W^-$}
\ArrowLine(260,260)(280,240)\ArrowLine(280,240)(300,260)
\Text(270,240)[r]{$\chi^{(n)}$}\Text(295,240)[l]{$\chi^{(m)}$}
\Photon(280,240)(280,220){3}{2}
\Text(240,265)[b]{$l$}\Text(320,265)[b]{$l'$}
\Text(282,220)[l]{$Z$}
\Text(280,200)[]{\bf (e)}

\ArrowLine(0,160)(20,160)\ArrowLine(60,160)(80,160)
\DashArrowLine(20,160)(60,160){3}\Text(40,167)[b]{$G^-$}
\ArrowLine(20,160)(40,140)\ArrowLine(40,140)(60,160)
\Text(30,140)[r]{$\chi^{(n)}$}\Text(55,140)[l]{$\chi^{(m)}$}
\Photon(40,140)(40,120){3}{2}
\Text(0,165)[b]{$l$}\Text(80,165)[b]{$l'$}
\Text(42,120)[l]{$Z$}
\Text(40,100)[]{\bf (f)}

\ArrowLine(120,160)(135,160)\ArrowLine(135,160)(150,160)
\ArrowLine(150,160)(180,160)\ArrowLine(180,160)(200,160)
\Text(120,165)[b]{$l$}\Text(142,165)[b]{$l$}
\Text(165,165)[b]{$\chi^{(n)}$}\Text(200,165)[b]{$l'$}
\Photon(135,160)(135,120){3}{4}
\Text(137,120)[l]{$Z,\gamma$}
\PhotonArc(165,160)(15,180,360){2}{5}\Text(165,135)[]{$W^-$}
\Text(160,100)[]{\bf (g)}

\ArrowLine(240,160)(255,160)\ArrowLine(255,160)(270,160)
\ArrowLine(270,160)(300,160)\ArrowLine(300,160)(320,160)
\Text(240,165)[b]{$l$}\Text(262,165)[b]{$l$}
\Text(285,165)[b]{$\chi^{(n)}$}\Text(320,165)[b]{$l'$}
\Photon(255,160)(255,120){3}{4}
\Text(257,120)[l]{$Z,\gamma$}
\DashArrowArc(285,160)(15,180,360){3}\Text(285,135)[]{$G^-$}
\Text(280,100)[]{\bf (h)}

\ArrowLine(0,60)(20,60)\ArrowLine(20,60)(50,60)
\ArrowLine(50,60)(65,60)\ArrowLine(65,60)(80,60)
\Text(0,65)[b]{$l$}\Text(57,65)[b]{$l'$}
\Text(35,65)[b]{$\chi^{(n)}$}\Text(80,65)[b]{$l'$}
\Photon(65,60)(65,20){3}{4}
\Text(67,20)[l]{$Z,\gamma$}
\PhotonArc(35,60)(15,180,360){2}{5}\Text(35,35)[]{$W^-$}
\Text(40,0)[]{\bf (i)}

\ArrowLine(120,60)(140,60)\ArrowLine(140,60)(170,60)
\ArrowLine(170,60)(185,60)\ArrowLine(185,60)(200,60)
\Text(120,65)[b]{$l$}\Text(177,65)[b]{$l'$}
\Text(155,65)[b]{$\chi^{(n)}$}\Text(200,65)[b]{$l'$}
\Photon(185,60)(185,20){3}{4}
\Text(187,20)[l]{$Z,\gamma$}
\DashArrowArc(155,60)(15,180,360){3}\Text(155,35)[]{$G^-$}
\Text(160,0)[]{\bf (j)}

\end{picture}\\[0.7cm]
\end{center}

\caption{Feynman graphs pertaining to the effective $\gamma ll'$ and
$Zll'$ couplings.}\label{fig:2}

\end{figure}
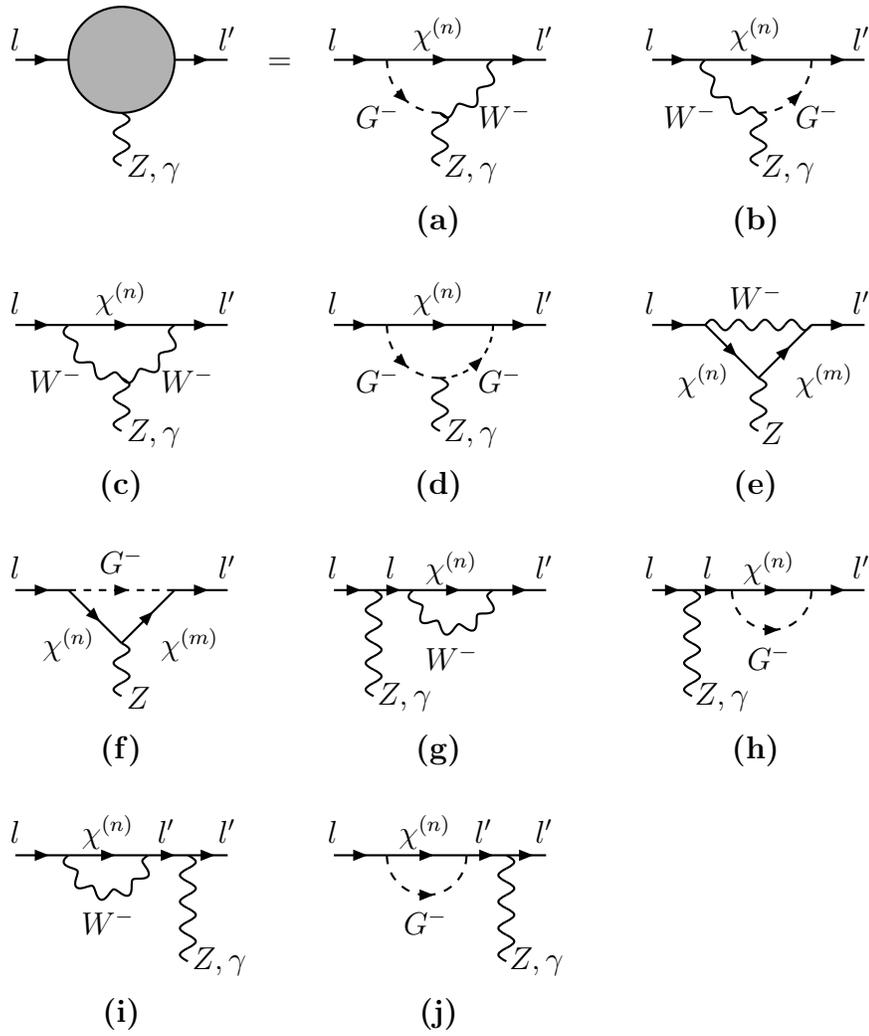

As  shown in  Fig.\ \ref{fig:2}, KK  neutrinos  can  give  rise to the
photonic FCNC decays  $\mu \to e \gamma$  and  $\tau \to e \gamma$  or
$\mu\gamma$. The transition element  for the generic decay $l(p_l) \to
l'(p_{l'}) \gamma (q)$ may conveniently be given by
\begin{eqnarray}
  \label{Tmuegamma} 
{\cal T} (l \to l' \gamma ) & =& \frac{ie\alpha_ws_w^2}{16\pi M_W^2}\ 
G_\gamma^{l l'}\, \varepsilon^\mu_\gamma\, \bar{u}_{l'}\, 
 i\sigma_{\mu\nu} q^\nu \Big[\, m_{l'} (1+\gamma_5) + 
 m_{l}(1-\gamma_5)\, \Big] u_l\, ,
\end{eqnarray}
where $G_\gamma^{l l'}$ is the composite form factor
\begin{equation}
  \label{Ggamma} 
G_\gamma^{l l'}\ =\ \sum^{\infty}_{n=-\infty}
B^\ast_{l,n} B_{l',n} G_\gamma(\lambda_n)\, ,
\end{equation}
with
\begin{equation}
  \label{Ggamma0}
G_\gamma (x) \ =\ -\, \frac{2x^3 + 5x^2 - x}{4 (1-x)^3}\ -\ 
\frac{3x^3\, \ln x}{2(1-x)^4}\ .
\end{equation}
{}From  this   very last  equation, it   is obvious that  the dominant
contribution to $G_\gamma^{l l'}$  comes from values of $\lambda_{(n)}
\gg 1$. Then, neglecting terms subleading in $\lambda_{(n)}$ and using
the fact that $G_\gamma  (x) = 1/2$ in the  infinite limit of $x$, the
composite form factor $G_\gamma^{l l'}$ may easily be evaluated to be
\begin{equation}
  \label{Ggamma1}
G_\gamma^{l l'}\ =\ \frac{1}{2}\; s_L^{\nu_l} s_L^{\nu_{l'}}\, ,
\end{equation}
where the   mixing   parameters, $s_L^{\nu_l}$, are  defined   in Eq.\
(\ref{snul})  and estimated  in Eq.\ (\ref{XiXi}). 

%******************************************************************
%   Figure 2
%******************************************************************

Employing  Eq.\  (\ref{Ggamma1}),  the branching ratio    of $l \to l'
\gamma$ may be determined by
\begin{equation}
  \label{Bllgamma}
B(l \to l' \gamma )\ =\ \frac{\alpha^3_w s^2_w}{256\pi^2}\, 
\frac{m^4_l}{M^4_W}\, \frac{m_l}{\Gamma_l}\, |G_\gamma^{l l'}|^2\
\approx\ \frac{\alpha^3_w s^2_w}{1024\pi^2}\, 
\frac{m^4_l}{M^4_W}\, \frac{m_l}{\Gamma_l}\; (s_L^{\nu_l})^2 
(s_L^{\nu_{l'}})^2\, .
\end{equation}
On  the  experimental side, we have \cite{PDG}
\begin{eqnarray}
  \label{Bexpllgamma}
B_{\rm exp} (\mu \to e \gamma) \!&<&\! 4.9\,(6.0) \times 10^{-11}\, ,\quad
B_{\rm exp} (\tau \to e \gamma)\ < \ 2.7\,(3.3)\times 10^{-6}\, ,\nonumber\\
B_{\rm exp} (\tau\to \mu \gamma) \!&<&\! 3.0\,(3.7)\times 10^{-6}\, ,
\end{eqnarray}
at the  90\% confidence  level  (CL).  The numbers in  the parentheses
refer  to upper limits at the  $2\sigma$ level.  Only the experimental
bound on $B(\mu  \to e \gamma)$ can  lead to stronger constraints than
those presented in Section 3.  Using the experimentally measured value
for the muon width, $\Gamma_\mu = 2.997\times 10^{-19}$ GeV, we obtain
the $2\sigma$ upper limit on the product $s_L^{\nu_e} s_L^{\nu_\mu}$:
\begin{equation}
  \label{limit1}
s_L^{\nu_e} s_L^{\nu_\mu}\  <\ 4.5\times 10^{-4}\, .
\end{equation}
If we take a scenario with $h_e = h_\mu = 1$  and $m=100$ GeV ($m=0$),
this very last bound translates into the lower limits on $M_F$:
\begin{equation}
  \label{mueMF}
M_F\ \stackrel{>}{{}_\sim}\ 75\ (165)\,,\ 43\,,\ 33\ {\rm TeV}\, ,
\end{equation}
for 2, 3 and 6 large compact dimensions,  respectively. This result is
in  qualitative  agreement with Ref.\   \cite{FP} for the scenario
with $m=0$.

In the   following,   we shall see  that,    owing  to the  cumulative
non-decoupling  effect of KK neutrinos,    the non-observation of  the
decay $\mu \to eee$ and  the absence of $\mu$-to-$e$-conversion events
in  nuclei can lead  to a much tighter bound   than that given by Eq.\ 
(\ref{mueMF}).

\subsection{Neutrinoless three-body  decays  of  the $\mu$ and $\tau$
leptons}

We shall calculate the non-decoupling loop effects  of KK neutrinos in
three-body  decays of  the $\mu$  and $\tau$ leptons:  $\mu  \to eee$,
$\tau  \to eee$, $\tau\to e\mu\mu$,  $\tau\to \mu\mu\mu$, and $\tau\to
ee\mu$.

%******************************************************************
%   Figure 3
%******************************************************************
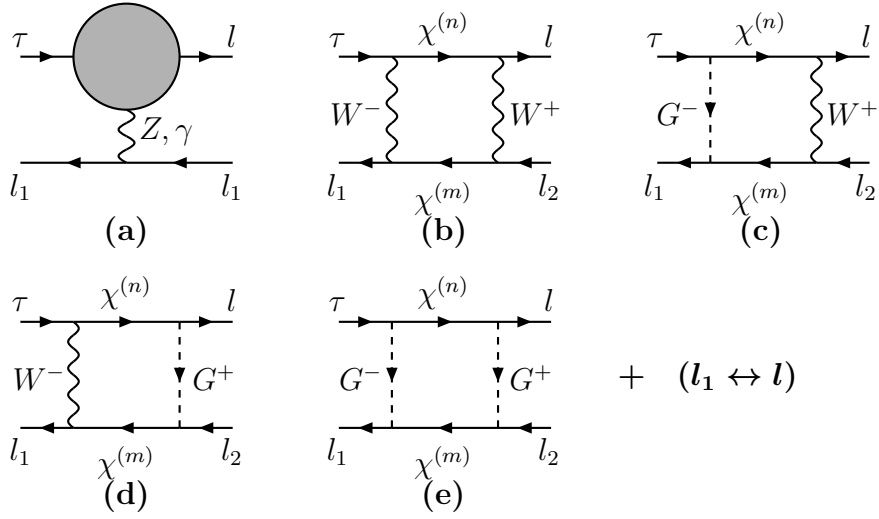
\begin{figure}

\begin{center}
\begin{picture}(360,200)(0,190)
\SetWidth{0.8}

\ArrowLine(0,360)(20,360)\ArrowLine(60,360)(80,360)
\GCirc(40,360){20}{0.7}\Photon(40,340)(40,320){3}{2}
\Text(0,365)[b]{$\tau$}\Text(80,365)[b]{$l$}
\Text(45,333)[l]{$Z,\gamma$}
\ArrowLine(40,320)(0,320)\ArrowLine(80,320)(40,320)
\Text(0,317)[t]{$l_1$}\Text(80,317)[t]{$l_1$}
\Text(40,295)[]{\bf (a)}

\ArrowLine(120,360)(140,360)\ArrowLine(180,360)(200,360)
\ArrowLine(140,360)(180,360)\Text(160,368)[b]{$\chi^{(n)}$}
\ArrowLine(140,320)(120,320)\ArrowLine(200,320)(180,320)
\ArrowLine(180,320)(140,320)\Text(160,314)[t]{$\chi^{(m)}$}
\Photon(140,360)(140,320){2}{4}\Photon(180,360)(180,320){2}{4}
\Text(120,365)[b]{$\tau$}\Text(200,365)[b]{$l$}
\Text(120,317)[t]{$l_1$}\Text(200,317)[t]{$l_2$}
\Text(137,340)[r]{$W^-$}\Text(185,340)[l]{$W^+$}
\Text(160,295)[]{\bf (b)}

\ArrowLine(240,360)(260,360)\ArrowLine(300,360)(320,360)
\ArrowLine(260,360)(300,360)\Text(280,368)[b]{$\chi^{(n)}$}
\ArrowLine(260,320)(240,320)\ArrowLine(320,320)(300,320)
\ArrowLine(300,320)(260,320)\Text(280,314)[t]{$\chi^{(m)}$}
\DashArrowLine(260,360)(260,320){3}\Photon(300,360)(300,320){2}{4}
\Text(240,365)[b]{$\tau$}\Text(320,365)[b]{$l$}
\Text(240,317)[t]{$l_1$}\Text(320,317)[t]{$l_2$}
\Text(257,340)[r]{$G^-$}\Text(305,340)[l]{$W^+$}
\Text(280,295)[]{\bf (c)}

\ArrowLine(0,260)(20,260)\ArrowLine(60,260)(80,260)
\ArrowLine(20,260)(60,260)\Text(40,268)[b]{$\chi^{(n)}$}
\ArrowLine(20,220)(0,220)\ArrowLine(80,220)(60,220)
\ArrowLine(60,220)(20,220)\Text(40,214)[t]{$\chi^{(m)}$}
\Photon(20,260)(20,220){2}{4}\DashArrowLine(60,260)(60,220){3}
\Text(0,265)[b]{$\tau$}\Text(80,265)[b]{$l$}
\Text(0,217)[t]{$l_1$}\Text(80,217)[t]{$l_2$}
\Text(17,240)[r]{$W^-$}\Text(65,240)[l]{$G^+$}
\Text(40,195)[]{\bf (d)}

\ArrowLine(120,260)(140,260)\ArrowLine(180,260)(200,260)
\ArrowLine(140,260)(180,260)\Text(160,268)[b]{$\chi^{(n)}$}
\ArrowLine(140,220)(120,220)\ArrowLine(200,220)(180,220)
\ArrowLine(180,220)(140,220)\Text(160,214)[t]{$\chi^{(m)}$}
\DashArrowLine(140,260)(140,220){3}\DashArrowLine(180,260)(180,220){3}
\Text(120,265)[b]{$\tau$}\Text(200,265)[b]{$l$}
\Text(120,217)[t]{$l_1$}\Text(200,217)[t]{$l_2$}
\Text(137,240)[r]{$G^-$}\Text(185,240)[l]{$G^+$}
\Text(160,195)[]{\bf (e)}

\Text(260,240)[]{\boldmath $+\quad ( l_1 \leftrightarrow l)$}

\end{picture}\\
\end{center}

\caption{Feynman graphs pertaining to the decay $\tau\to
  ll_1\bar{l}_2$.}\label{fig:3}

\end{figure}                                  

For convenience, we  first  consider the  generic decay process  $\tau
(p_\tau)  \to  l (p_l) l_1   (p_{l_1})  \bar{l}_2 (p_{l_2})$, which is
induced  by the  loop    graphs  shown  in  Figs.\  \ref{fig:2}    and
\ref{fig:3}; analytic   expressions  for $\mu  \to   eee$ may  then be
obtained by   an obvious interchange   of the  $\tau$--$\mu$ kinematic
parameters.   As can be seen from  Fig.\ \ref{fig:3},  there are three
amplitudes  that contribute to the decay  $\tau \to  l l_1 \bar{l}_2$:
the photon-exchange     amplitude, ${\cal T}_\gamma$;    the $Z$-boson
mediated one,  ${\cal T}_Z$,  and  an  amplitude  describing  the  box
contribution, ${\cal T}_{\rm box}$.  The three contributing amplitudes
are given by
\begin{eqnarray}
  \label{Tgamma} 
{\cal T}_\gamma (\tau\to l l_1
\bar{l}_2)&=& -\frac{i\alpha_w^2s_w^2}{4M_W^2} \delta_{l_1 l_2}
\bar{u}_{l_1}\gamma^\mu v_{l_2}\ \bar{u}_{l} \Big\{\, F^{\tau l}_\gamma
\Big(\, \gamma_\mu\, -\, \frac{q_\mu \not\! q}{q^2}\Big)\, 
(1-\gamma_5) \nonumber\\ 
& & -iG_\gamma^{\tau l} \sigma_{\mu\nu}\frac{q^\nu}{q^2}
\Big[\,m_\tau (1+\gamma_5)\, +\, 
               m_{l}(1-\gamma_5)\,\Big]\,\Big\}\, u_\tau\, ,\\ 
  \label{TZ}
{\cal T}_Z(\tau\to l l_1 \bar{l}_2)&=&-\frac{i\alpha_w^2}{16M_W^2}\
\delta_{l_1 l_2} F_Z^{\tau l} \bar{u}_{l}\gamma_\mu(1-\gamma_5)u_\tau\
\bar{u}_{l_1}\gamma^\mu (1-4s_w^2-\gamma_5)v_{l_2} \, ,\\
   \label{Tbox}
{\cal T}_{\rm box}(\tau \to l l_1 \bar{l}_2)&=&-\frac{i\alpha_w^2}{16M_W^2}\
F_{\rm box}^{\tau ll_1l_2}\ \bar{u}_{l}\gamma_\mu(1-\gamma_5)u_\tau\
\bar{u}_{l_1}\gamma^\mu(1-\gamma_5)v_{l_2}\ , 
\end{eqnarray}  
where $q = p_\tau - p_l$ and \cite{IP}
\begin{eqnarray} 
  \label{Fgamma}
F_\gamma^{\tau l} &=& \sum^\infty_{n=-\infty} 
B^\ast_{\tau,n}B_{l,n} F_\gamma (\lambda_n)\ ,\\ 
  \label{FZ}
F_Z^{\tau l} &=&  \sum^\infty_{n,m = -\infty} 
B^\ast_{\tau,n} B_{l,m} \Big\{\, \delta_{nm}\,
\Big[\,  F_Z (\lambda_n)\, +\, 
                           2G_Z (0,\lambda_n)\, \Big]\nonumber\\
&&+\, C^\ast_{n,m}\, \Big[\, G_Z(\lambda_n,\lambda_m)\,
-\, G_Z(0,\lambda_n)\, -\, G_Z(0,\lambda_m)\, \Big]\, \Big\},\\ 
  \label{Fbox}
F_{\rm box}^{\tau l l_1 l_2} &=& \sum^\infty_{n,m = -\infty}
\Big\{\, \delta_{nm} \Big( B^\ast_{\tau, n} B_{l,m}\delta_{l_1l_2}\, +\,
B^\ast_{\tau, n} B_{l_1,m}\delta_{ll_2}\Big)\,
\Big[\, F_{\rm box}(0,\lambda_n)\, -\, F_{\rm box}(0,0)\, \Big]\nonumber\\
&&+\, B^\ast_{\tau,n} B^\ast_{l_2,m}\, \Big(\,B_{l,n}B_{l_1,m}\, +\, 
B_{l_1,n}B_{l,m}\,\Big)\, \Big[\, F_{\rm box}(\lambda_n,\lambda_m)\nonumber\\
&&-\, F_{\rm box}(0,\lambda_n)\, -\, F_{\rm box}(0,\lambda_m)\, +\,
F_{\rm box}(0,0)\, \Big]\, \Big\}\, .
\end{eqnarray}
Note that the composite form factor  $G_\gamma^{\tau l}$ is defined in
Eq.\ (\ref{Ggamma}).  In deriving  the expressions in Eqs.\ (\ref{FZ})
and (\ref{Fbox}), we made use  of  the unitarity relations, which  are
obeyed by the $B$ and $C$ matrices  defined in Eqs.\ (\ref{Bmatr}) and
(\ref{Cmatr}).  Furthermore, the  analytic forms of the loop functions
$F_\gamma$, $F_Z$, $G_Z$, and $F_{\rm  box}$, which may also be  found
in \cite{IP,LI}, are given by
\begin{eqnarray}
  \label{Fgamma0}
F_\gamma (x) &=& \frac{ 7x^3 - x^2 - 12 x}{12 (1 -x)^3}\, -\,
\frac{(x^4 - 10x^3 +12 x^2 )\ln x}{6 (1 -x )^4}\ ,\\
  \label{FZ0}
F_Z (x) & =& -\, \frac{5x}{2 (1-x)}\, -\, \frac{5x^2\, \ln x}{2 (1-x^2)}\ ,\\
  \label{GZ0}
G_Z (x,y) &=& -\, \frac{1}{2 (x-y)}\, \bigg[\, \frac{x^2 (1-y) \ln x}{1-x}\, 
-\, \frac{y^2 (1-x) \ln y}{1-y}\, \bigg]\, ,\\
  \label{Fbox0}
F_{\rm box} (x,y) &=& 
\frac{1}{x-y}\, \bigg[\, \bigg( 1\, +\, \frac{xy}{4}\bigg)
\bigg( \frac{1}{1-x}\, +\, \frac{x^2\ln x}{(1-x)^2}\, -\, \frac{1}{1-y}\, 
-\, \frac{y^2 \ln y}{(1-y)^2}\bigg)\nonumber\\
&&-\, 2xy\, \bigg( \frac{1}{1-x}\, +\, \frac{x \ln x}{(1-x)^2}\, -\,
\frac{1}{1-y}\, -\, \frac{y \ln y}{(1-y)^2}\, \bigg)\, \bigg]\, .
\end{eqnarray}
As  we discussed  in Section 5,   to  leading order  in $M_F/M_W$, the
dominant  contributions to   the    composite form  factors  in  Eqs.\
(\ref{Fgamma})--(\ref{Fbox}) come from KK  states heavier than the $W$
boson. More explicitly, we have
\begin{eqnarray}
  \label{Fgamma1}
F_\gamma^{\tau l} &\approx & -\,\frac{7}{12}\, s_L^{\nu_\tau} s_L^{\nu_{l}}\,
 -\, \frac{c_\delta}{6}\, s_L^{\nu_\tau} s_L^{\nu_{l}} \ln\lambda_F\, , \\  
  \label{FZ1}
F_Z^{\tau l}&\approx & 
\bigg(\, \frac{5}{2}\, -\, \frac{3 c_\delta}{2}\, \ln\lambda_F \bigg)
s_L^{\nu_\tau}s_L^{\nu_{l}}\,
-\, \frac{d_\delta}{2} s_L^{\nu_\tau}s_L^{\nu_{l}}\sum_{k=e,\mu,\tau}\ 
(s_L^{\nu_k})^2 \lambda_F\, ,\\ 
  \label{Fbox1}
F_{\rm box}^{\tau ll_1l_2}&\approx & -\,
s_L^{\nu_\tau}s_L^{\nu_{l}}\delta_{l_1l_2}\, -\, 
s_L^{\nu_\tau}s_L^{\nu_{l_1}}\delta_{ll_2}\, +\,
\frac{d_\delta}{2}s_L^{\nu_\tau}s_L^{\nu_{l}}s_L^{\nu_{l_1}}
s_L^{\nu_{l_2}}\, \lambda_F\, , 
\end{eqnarray}
where $\lambda_F = M^2_F/M^2_W$,  $c_\delta = 1/2$,  for $\delta = 2$,
and   $c_\delta   = 1$,  for  $\delta  >2$.  In Eqs.\  (\ref{FZ1}) and
(\ref{Fbox1}), $d_\delta$ are dimension-dependent correction  factors,
which are given in Eq.\ (\ref{ddel}).

The branching ratios for the decays $\tau  \to e \mu \mu$ and $\tau\to
eee$  were  calculated in \cite{IP,MPLAP}. Their  analytic expressions
may be cast into the form:
\begin{eqnarray}
  \label{Btemumu}
B(\tau \to e \mu \mu ) &=& 
     \frac{\alpha_w^4}{24576\pi^3}\ \frac{m_\tau^4}{M_W^4}\
     \frac{m_\tau}{\Gamma_\tau}\ \bigg\{\, 
   |F_{\rm box}^{\tau e \mu\mu }+F_Z^{\tau e}
            -2s_w^2(F_Z^{\tau e}-F_\gamma^{\tau e})|^2 \nonumber\\
&&+\, 4s_w^4 |F_Z^{\tau e}-F_\gamma^{\tau e}|^2 
  +\ 8s_w^2\: {\rm Re} \Big[ (F_Z^{\tau e}+F_{\rm box}^{\tau e\mu\mu})
     G_\gamma^{\tau e\:\ast}\Big]\nonumber\\
&& -\, 32s_w^4\: 
{\rm Re} \Big[(F_Z^{\tau e}-F_\gamma^{\tau e})G_\gamma^{\tau e\:\ast}\Big] 
  \,+\, 32s_w^4 |G_\gamma^{\tau e}|^2\:
     \bigg( \ln\frac{m_\tau^2}{m_\mu^2}-3\bigg)\, \bigg\}\,,\\
  \label{Bteee}
B(\tau \to eee ) &=& 
     \frac{\alpha_w^4}{24576\pi^3}\ \frac{m_\tau^4}{M_W^4}\
     \frac{m_\tau}{\Gamma_\tau}\bigg\{\, 2 |{\textstyle \frac{1}{2}}
        F_{\rm box}^{\tau eee}+F_Z^{\tau e}
            -2s_w^2(F_Z^{\tau e}-F_\gamma^{\tau e})|^2\nonumber\\
&&  +\, 4s_w^4 |F_Z^{\tau e}-F_\gamma^{\tau e}|^2\, +\, 
16s_w^2\: {\rm Re} \Big[(F_Z^{\tau e}+{\textstyle \frac{1}{2}}
     F_{\rm box}^{\tau eee})
      G_\gamma^{\tau e\:\ast}\Big]\nonumber\\
&&-\, 48s_w^4\: {\rm Re} \Big[\,(F_Z^{\tau e}-F_\gamma^{\tau e})
G_\gamma^{\tau e\:\ast}\,\Big]\, +\, 32s_w^4 |G_\gamma^{\tau e} |^2\: 
     \bigg(\ln\frac{m_\tau^2}{m_e^2}-\frac{11}{4}\bigg)\, \bigg\}\, .\qquad
\end{eqnarray}
The branching ratio  for the decay $\mu   \to eee$ may be obtained  by
making, in Eq.\  (\ref{Bteee}), the obvious  replacements: $m_\tau \to
m_\mu$ and $m_\mu \to m_e$.   Taking the dominant non-decoupling parts
of  the composite form   factors  into account,   we  obtain  for  the
branching ratios:
\begin{eqnarray}
  \label{Btauemm}
B(\tau \to e \mu \mu)
&\simeq & \frac{\alpha_w^4}{98304\pi^3}\ \frac{m_\tau^4}{M_W^4}\
\frac{m_\tau}{\Gamma_\tau} \frac{M^4_F}{M^4_W}\ d^2_\delta\,
(s_L^{\nu_\tau})^2(s_L^{\nu_{e}})^2 \Bigg\{\, (s_L^{\nu_{\mu}})^4\nonumber\\ 
&& +\ 2(1-2s^2_w)(s_L^{\nu_\mu})^2 
\Big[\sum_{l=e,\mu,\tau} (s_L^{\nu_l})^2\Big] + 8s^4_w
\Big[\sum_{l=e,\mu,\tau} (s_L^{\nu_l})^2\Big]^2\ \Bigg\}\, ,\\  
  \label{Btaueee}
B (\tau\to e e e)
&\simeq & \frac{\alpha_w^4}{98304\pi^3}\ \frac{m_\tau^4}{M_W^4}\
\frac{m_\tau}{\Gamma_\tau} \frac{M^4_F}{M^4_W}\ d^2_\delta\,
(s_L^{\nu_\tau})^2 (s_L^{\nu_e})^2 
\Bigg\{\, \frac{1}{2}(s_L^{\nu_e})^4\nonumber\\ 
&& +\ 2(1-2s^2_w)(s_L^{\nu_e})^2 
\Big[\sum_{l=e,\mu,\tau} (s_L^{\nu_l})^2\Big] + 12s^4_w
\Big[\sum_{l = e,\mu,\tau} (s_L^{\nu_l})^2\Big]^2\ \Bigg\}\, .  
\end{eqnarray} 

The present experimental upper limits  on the branching ratios for the
lepton-flavour-violating three-body decays of $\mu$ and $\tau$ leptons
are \cite{PDG}
\begin{eqnarray}
  \label{Bexp3lll}
B_{\rm exp} (\mu \to eee) &<& 1.0\, (1.2)\times 10^{-12}\,,\qquad
B_{\rm exp} (\tau \to e\mu\mu)\ <\ 1.8\, (2.2)\times 10^{-6}\,,\nonumber\\
B_{\rm exp} (\tau \to eee) &<& 2.9\, (3.5)\times 10^{-6}\,,\qquad\ 
B_{\rm exp} (\tau \to \mu ee)\ <\ 1.5\, (1.8)\times 10^{-6}\,,\nonumber\\
B_{\rm exp} (\tau \to \mu\mu\mu) &<& 1.9\, (2.3)\times 10^{-6}\, ,
\end{eqnarray}
at   the  90\% confidence ($2\sigma$)    level.  In the scenario, with
complete Yukawa-coupling universality, i.e.\  $h_e = h_\mu =  h_\tau =
h$, the most  severe bound on $M_F$ comes  from $B_{\rm  exp} (\mu \to
eee)$. For $h = 1$ and $m = 100$ GeV ($m=0$), we obtain the lower limits:
\begin{equation}
  \label{mueeeMF}
M_F\ \stackrel{>}{{}_\sim}\ 
\left\{ \begin{array}{l} 
250\ (1000)\  {\rm TeV}\,,\quad {\rm for}\ \delta=2 \\ 
210\  {\rm TeV}\,,\quad {\rm for}\ \delta = 3-6.
\end{array}\ \right.
\end{equation}
If    the limits in    Eq.\  (\ref{mueeeMF}) are  implemented in  this
scenario,  the predictions obtained for  $B(\tau \to eee)$ and $B(\tau
\to e\mu\mu)$ are hopelessly small, of  order $10^{-12}$, to be tested
in any future experiment.

The high upper   bounds  on $M_F$ given   in  Eqs.\ (\ref{mueMF})  and
(\ref{mueeeMF})  may  be  completely avoided   in  a scenario in which
$h_\mu =   0$ and $h_e  = h_\tau  = h$.   In this case,  $M_F$ is only
constrained by  limits  presented in  Table  \ref{Tab1}.  Implementing
these  constraints  for  $h=1$ and   $m   = 100$ GeV  ($m=0$)
(see  also  Eq.\ (\ref{MFII})), we find
\begin{eqnarray}
  \label{Bthlll}
B (\tau \to eee) &\stackrel{<}{{}_\sim}& 
               7.2\times 10^{-11}\ (1.4\times 10^{-10}),\ 
6.7\times 10^{-10}\, ,\nonumber\\
B (\tau \to e\mu\mu) &\stackrel{<}{{}_\sim}& 
               5.9\times 10^{-11}\ (1.2\times 10^{-10}),\ 
6.8\times 10^{-10}\, .\qquad
\end{eqnarray}
The numbers in the parentheses correspond to  numerical estimates in a
theory  with  $\delta  =   2,\ 3    -   6$  large extra    dimensions,
respectively. The predictions given in Eq.\ (\ref{Bthlll}) for $B(\tau
\to  e\mu\mu )$ and  $B(\tau \to eee  )$  may be enhanced  by a factor
$10^3$ for  large Yukawa couplings $h\sim 5$,  to the $10^{-7}$ level,
which might be probed in future experiments.

\subsection{Coherent $\mu$-to-$e$ conversion in nuclei} 

Coherent  $\mu   \to   e$   conversion  in   nuclei,   e.g.\   $\mu^-\
{}^{48}_{22}{\rm Ti}  \to e^-\  {}^{48}_{22}{\rm Ti}$, constitutes one
of the   most sensitive experiments    that can set severe  bounds  on
lepton-flavour-violating physics  \cite{FW,JDV}.  The existence  of KK
neutrinos  may give rise  to  a  sizeable $\mu \to  e$  conversion  in
nuclei, which   comes  from    the  following $\gamma$-,    $Z$-   and
box-mediated transition amplitudes:
\begin{eqnarray}
  \label{mueTg}
{\cal T}_\gamma (\mu \to e) &=& \frac{i\alpha_w^2s_w^2}{4M_W^2}\ 
\bar{u}_e \bigg[\, F^{\mu e}_\gamma \gamma_\nu (1-\gamma_5)\, 
-\, iG_\gamma^{\mu e} \sigma_{\nu\lambda}\frac{q^\lambda}{q^2} m_\mu 
(1+\gamma_5)\,\bigg]\, u_\mu \nonumber\\
&&\times\, \bigg(\, \frac{2}{3} \bar{u}_u \gamma^\nu u_u\, -\, 
\frac{1}{3} \bar{u}_d \gamma^\nu u_d\,\bigg)\, ,\\
  \label{mueTZ}
{\cal T}_Z(\mu \to e)&=& \frac{i\alpha_w^2}{16M_W^2}\
F_Z^{\mu e}\, \bar{u}_e \gamma_\nu (1-\gamma_5) u_\mu\,
\bigg[\, \bigg(\, 1 - \frac{8}{3}\, s^2_w\, \bigg)\,
 \bar{u}_u\gamma^\nu u_u \nonumber\\
&&-\, \bigg(\, 1 - \frac{4}{3}\, s^2_w\, \bigg)\, 
\bar{u}_d\gamma^\nu u_d\,\bigg]\, ,\\
   \label{mueTbox}
{\cal T}_{\rm box}(\mu \to e)&=& -\, \frac{i\alpha_w^2}{16M_W^2}\
\bar{u}_e \gamma_\nu (1-\gamma_5) u_\mu\, \Big[
F_{\rm box}^{\mu e uu}\ \bar{u}_u\gamma^\nu u_u\, +\,
F_{\rm box}^{\mu e dd}\ \bar{u}_d\gamma^\nu u_d\,\Big]\, , 
\end{eqnarray}
where
\begin{eqnarray}
  \label{Fboxuu}
F_{\rm box}^{\mu euu} &=& \sum^\infty_{n = -\infty}
B^\ast_{\mu, n} B_{e,n}\, \Big[\, F_{\rm box}(0,\lambda_n)\, 
-\, F_{\rm box}(0,0)\, \Big]\, ,\nonumber\\
  \label{Fboxdd}
F_{\rm box}^{\mu e dd} &=& \sum^\infty_{n = -\infty} \Big\{\,
B^\ast_{\mu, n} B_{e,n}\, \Big[\, F_{\rm box}(0,\lambda_n)\, 
-\, F_{\rm box}(0,0)\, \Big]\, +\,
B^\ast_{\mu, n} B_{e,n} |V_{td}|^2\, \Big[\, 
F_{\rm box}(\lambda_t,\lambda_n)\nonumber\\
&&-\, F_{\rm box}(0,\lambda_n)\, -\, F_{\rm box}(0,\lambda_t)\, 
+\, F_{\rm box}(0,0)\, \Big]\, \Big\}\, ,
\end{eqnarray}
with $\lambda_t = m^2_t/M^2_W$ and $|V_{td}| \approx 0.01$ \cite{PDG}.
In Eqs.\  (\ref{mueTg})--(\ref{mueTbox}), we have  only considered the
vectorial coupling  of  the $u$  and $d$  quarks, which is  coherently
enhanced in  nuclei, whereas  the axial  part  of the quark  couplings
describes spin-dependent  interactions,  and therefore,   their  total
contribution is almost vanishing \cite{FW,JDV}.

Following  \cite{FW,JDV}, we  consider the  kinematic  approximations:
$q^2 \approx -m^2_\mu$ and  $p^0_e \approx |\vec{p}_e| \approx m_\mu$,
which are  applicable for $\mu \to  e$ conversion in nuclei. Then, for
nuclei with nucleon numbers $(N,Z)$, we obtain
\begin{equation}
  \label{Bmueconv}
B_{\mu e} (N,Z)\ \equiv\ \frac{\Gamma [\mu\, (N,Z)\to e\, (N,Z) ] }{
\Gamma [ \mu\, (N,Z) \to {\rm capture}]}\
\approx\ \frac{ \alpha^3_{\rm em} \alpha^4_w m^5_\mu}{32\pi^2
M^4_W \Gamma_{\rm capt.} }\, \frac{Z^4_{\rm eff}}{Z}\,
|F(-m^2_\mu)|^2\, |Q_W|^2\, ,
\end{equation}
where $\alpha_{\rm em} = 1/137$ is  the electromagnetic fine structure
constant, $Z_{\rm eff}$ is the   effective atomic number of  coherence
(e.g.\  $Z_{\rm   eff}  \approx  17.6$   for  ${}^{48}_{22}{\rm  Ti}$)
\cite{Zeff}, $|F(-m^2_\mu)| \approx 0.54$  is the nuclear form  factor
\cite{FPap}, and $Q_W = V_u (2Z +N) + V_d (Z+2N)$, with
\begin{eqnarray}
  \label{Vu}
V_u &=& \frac{2}{3}\, s^2_w\, \Big(\, F^{\mu e}_\gamma\, -\, 
G^{\mu e}_\gamma\, -\, F^{\mu e}_Z\, \Big)\, +\, \frac{1}{4}\, 
\Big(\, F^{\mu e}_Z\, -\, F^{\mu e uu}_{\rm box}\, \Big)\, ,\\
  \label{Vd}
V_d &=& -\, \frac{1}{3}\, s^2_w\, 
\Big(\, F^{\mu e}_\gamma\, -\, G^{\mu e}_\gamma
\, -\, F^{\mu e}_Z\, \Big)\, -\, \frac{1}{4}\, \Big(\, F^{\mu e}_Z\, +\,
F^{\mu e dd}_{\rm box}\, \Big)\, .
\end{eqnarray}
The  strictest upper  limit on   $B_{\mu e}  (N,Z)$ is obtained   from
experimental data of $\mu \to e$  conversion in ${}^{48}_{22}{\rm Ti}$
\cite{SINDRUM}:
\begin{equation}
  \label{mueconvexp}
B^{\rm exp}_{\mu  e} (26,22)\ <\ 4.3\, (5.24) \times 10^{-12}\, , 
\end{equation}
at the 90\%  confidence $(2\sigma)$ level.  For numerical predictions,
we  use the experimental   value  of the  muon nuclear   capture rate,
$\Gamma  [ \mu\ {}^{48}_{22}{\rm    Ti}  \to {\rm  capture}]   \approx
1.705\times 10^{-18}$ GeV \cite{SMR}.

For definiteness,  we shall now consider  the scenario with all Yukawa
couplings equal  to   unity and  $m =  100$  GeV  ($m=0$).  Then,  the
experimental  upper bound on $B_{\mu  e}  (26,22)$ gives the following
lower limits on $M_F$:
\begin{equation}
  \label{BmueMF}
M_F\ \stackrel{>}{{}_\sim}\ 
\left\{ \begin{array}{l} 
380\ (1430)\   {\rm TeV}\,,\quad {\rm for}\ \delta=2 \\ 
310\  {\rm TeV}\,,\quad {\rm for}\ \delta = 3-6.
\end{array}\ \right.
\end{equation}
These limits represent  the most severe bounds that  can be derived on
the   fundamental  quantum   gravity  scale   $M_F$   from  laboratory
experiments, for large higher-dimensional Yukawa couplings, e.g.\ $h_l
\sim 1$.

\subsection{Lepton-flavour violation and breaking of universality at the
$Z$ peak}

Loop effects of KK neutrinos may induce a number of lepton-flavour and
universality-breaking phenomena  on  the  $Z$-boson  pole.   The  most
striking  experimental signals of new physics  would be FCNC $Z$-boson
decays into different leptons,  e.g.\ $Z\to ll'$.  Further new-physics
signals would  be the  detection  of breaking  of universality  in the
leptonic partial widths $Z\to  l\bar{l}$ and in the $Z$-boson leptonic
asymmetries, which  may  be  probed  by  looking  at the   observables
\cite{BKPS,BP}:
\begin{eqnarray}
  \label{Ubr}
U^{(ll')}_{\rm br} &=& \frac{\Gamma (Z \to l\bar{l} )\, -\, 
\Gamma ( Z\to l'\bar{l}' ) }{ \Gamma (Z \to l\bar{l} )\, +\, 
\Gamma ( Z\to l'\bar{l}' )}\, -\, U^{\rm PS}_{\rm br}\ =\
\frac{|{\cal T} (Z \to l\bar{l})|^2\, -\, |{\cal T} (Z \to l'\bar{l}')|^2}
{|{\cal T} (Z \to l\bar{l})|^2\, +\, |{\cal T} (Z \to l'\bar{l}')|^2 }\, 
,\quad\\
  \label{DAbr} 
\Delta {\cal A}_{ll'} &=& \frac{{\cal A}_l\, -\, {\cal A}_{l'}}{
{\cal A}_l\, +\, {\cal A}_{l'}}\ =\ 
\bigg( \frac{1}{{\cal A}^{\rm SM}_l}\, -\, 1\,\bigg)\, U^{(ll')}_{\rm br}\, ,
\quad {\rm with}\quad l \neq l'\, ,
\end{eqnarray}
where $U^{\rm PS}_{\rm br}$   indicates known phase-space  corrections
due to the finiteness of the charged-lepton masses, and ${\cal A}^{\rm
  SM}_l = 0.14$ is the SM prediction  \cite{PDG}. In deriving the last
equality of  Eq.\ (\ref{DAbr}), we  used the theoretical fact that the
induced coupling of the $Z$ boson to  charged leptons is predominantly
left-handed, as is the case in our higher-dimensional singlet-neutrino
model (cf.\ Eq.\ (\ref{TZ})).

The branching ratio of the FCNC decay $Z\to l l'$ is given by
\begin{eqnarray}
  \label{BZetau}
B(Z\to l \bar{l}'\ {\rm or}\ \bar{l} l' ) &=&
\frac{\alpha_w^3}{48\pi^2c_w^2} \frac{M_Z}{\Gamma_Z} 
                                 |{\cal F}_Z^{ll'}(M^2_Z)|^2 \nonumber\\
&\approx & \frac{\alpha_w^3}{768\pi^2c_w^2} \frac{M_Z}{\Gamma_Z}
\frac{M^4_F}{M^4_W}\, d^2_\delta\, (s_L^{\nu_l})^2 (s_L^{\nu_{l'}})^2
\Big[ \sum_{k=e,\mu,\tau} (s_L^{\nu_k})^2 \Big]^2\, ,
\end{eqnarray}
where  the loop function  ${\cal F}_Z^{ll'}(M^2_Z)$  was calculated in
\cite{KPS,IP}. To obtain the  last equality in Eq.\ (\ref{BZetau}), we
used   the  approximation:  ${\cal  F}_Z^{ll'}(M^2_Z)   \approx  {\cal
F}_Z^{ll'}(0)  = F_Z^{ll'}/2$,  where   $F_Z^{ll'}$ is  given by  Eq.\
(\ref{FZ}).     Furthermore,   the   universality-breaking  observable
$U^{(ll')}_{\rm br}$ is found to be
\begin{eqnarray}
  \label{Ubrth}
U^{(ll')}_{\rm br} & =& \frac{\alpha_w}{2\pi}\, 
\frac{ 1 - 2s^2_w }{ (1-2s^2_w)^2 + 4 s^4_w}\ \Big[\, 
{\cal F}^{ll}_Z (M^2_Z)\, -\, {\cal F}^{l'l'}_Z (M^2_Z)\, \Big]\nonumber\\
&\approx & \frac{\alpha_w}{8\pi}\, 
\frac{ 1 - 2s^2_w }{ (1-2s^2_w)^2 + 4 s^4_w}\ \frac{M^2_F}{M^2_W}\,
d_\delta\, \Big[\,(s_L^{\nu_l})^2\, 
-\, (s_L^{\nu_{l'}})^2\,\Big]\, \Big[
\sum_{k=e,\mu,\tau} (s_L^{\nu_k})^2 \Big]\, .
\end{eqnarray}

The current     experimental    situation  on    $B(Z\to   ll')$   and
$U^{(ll')}_{\rm br}$ is as follows \cite{PDG}:
\begin{equation}
  \label{BZll'}
B_{\rm exp}( Z\to e\mu )\, <\, 1.7\times 10^{-6}\, ,\quad 
B_{\rm exp}( Z\to e\tau )\, <\, 9.8\times 10^{-6}\, , \quad 
B_{\rm exp}( Z\to \mu\tau )\, <\, 1.2\times 10^{-5}\, ,
\end{equation}
and, almost independently of the lepton flavours $l$ and $l'$,
\begin{equation} 
  \label{Ubrexp}
U^{(ll'){\rm exp}}_{\rm br}\ =\ \frac{1}{2}\, 
\bigg(1- \frac{\Gamma(Z\to l'l')}{\Gamma (Z\to ll)}\bigg)\ =\ 
5.0\times 10^{-3}\, , 
\end{equation}
at   the  $2\sigma$   level.    The     experimental limits on     the
universality-breaking parameter  $\Delta {\cal A}_{ll'}$  are slightly
weaker than  those derived by  $U^{(ll')}_{\rm br}$,  if one uses  the
relation  given by Eq.\  (\ref{DAbr}),  i.e.\ $\Delta {\cal A}_{ll'} <
3.0 \times 10^{-2}$.

In view of the  constraints derived in Sections 5.1--5.3,  new-physics
phenomena in  $Z$-boson decays  with  electrons and  muons only in the
final state are   far beyond the  realm of   detection.  Therefore, we
shall  discuss  the   numerical predictions  obtained    for the  FCNC
$Z$-boson decay $Z\to   e  \tau$ and $U^{(\mu  \tau)}_{\rm   br}$ in a
phenomenologically more  favourable scenario, with $h_e  = h_\tau = h$
and $h_\mu = 0$.  Considering $h = 1$ and $m = 100$ GeV ($m=0$), and the lower
limits on  $M_F$ stated in Eq.\ (\ref{MFII}),  we obtain the following
upper limits for $\delta = 2$ and 3--6:
\begin{eqnarray}
  \label{BthZll'}
B (Z \to e \tau) &\stackrel{<}{{}_\sim}& 
  2.9\times 10^{-10}\ (5.5\times 10^{-10}),\
3.8\times 10^{-9}\, ,\nonumber\\
   \label{Uthbr}
U_{\rm br}^{(\mu \tau)}  &\stackrel{<}{{}_\sim}& 
     5.0\times 10^{-5}\ (8.8\times 10^{-5}),\ 1.9\times 10^{-4}\, .\qquad
\end{eqnarray}
For maximal values  of  the Yukawa  couplings,  $h\sim 5$, the   above
predictions  for $B  (Z \to e   \tau)$ and  $U_{\rm  br}^{(\mu \tau)}$
increase by a factor $10^3$ and 25, respectively.

Finally,  it is illuminating to discuss  the  upper bounds that can be
obtained for $M_F$ in a more weakly constrained scenario, i.e.\ $h_e =
h_\mu  = 0$ and $h_\tau  \not= 0$.  In  addition,  we assume that this
scenario  is only constrained by experimental  limits on the invisible
width of the $Z$ boson (cf.\ Eq.\ (\ref{MFIII})).  If $h_\tau = 1$ and
$m = 100$ GeV ($m=0$),  the experimental bounds on $U^{(e  \tau)}_{\rm
br}$ and $U^{(\mu \tau )}_{\rm br}$ yield
\begin{equation}
  \label{UbrMF}
M_F\ \stackrel{>}{{}_\sim}\ 
3\ (31)\  {\rm TeV}\,,\quad {\rm for}\ \delta = 2-6\, .
\end{equation}
Note that these  limits are  almost  comparable  to those deduced   by
experimental data  on     the  invisible $Z$-boson    width   in  Eq.\ 
(\ref{MFIII}).

\subsection{Contribution to the electroweak oblique parameters}

Finally, KK neutrinos may manifest their presence by inducing sizeable
contributions to the electroweak  oblique parameters $S$, $T$  and $U$
\cite{STU}.   We find that  a  cumulative non-decoupling  effect of KK
states also occurs here, very much like the previous flavour-dependent
observables.  For the kind of the  new physics we are considering, $T$
turns out   to be the  most  sensitive electroweak  oblique parameter,
which is related to Veltman's $\rho$ parameter \cite{Velt} through
\begin{equation}
  \label{rho}
\rho - 1\ =\ \alpha_{\rm em}(M_Z)\, T\ 
=\ \frac{\Pi^{\rm KK}_{WW} (0)}{M^2_W}\, 
-\, \frac{\Pi^{\rm KK}_{ZZ} (0)}{M^2_Z}\ .
\end{equation}
In  Eq.\ (\ref{rho}), $\Pi^{\rm KK}_{WW}   (0)$ and $\Pi^{\rm KK}_{ZZ}
(0)$ indicate the KK-neutrino  contributions to the $W$- and $Z$-boson
self-energies, respectively.

The most recent experimental constraint on $T$ is $T = -0.21 \pm 0.16\,
(+0.10)$,   for  $M_H  = M_Z\   (300\  {\rm  GeV})$ \cite{PDG}.   This
corresponds to the $2\sigma$ upper bound on $(\rho - 1)$:
\begin{equation}
  \label{rhoexp}
(\rho - 1)_{\rm exp}\ \le\ 2.42\times 10^{-3}\, ,
\end{equation}
for a Higgs-boson mass $M_H = 300$ GeV.  On the other hand, to leading
order in $M_F/M_W$, the electroweak  oblique parameter $(\rho -1)$  is
found to be
\begin{equation}
  \label{rhoth}
\rho - 1\ \approx\ \frac{\alpha_w}{16\pi}\, \frac{M^2_F}{M^2_W}\,
d_\delta\,\Big[\sum_{l=e,\mu,\tau} (s_L^{\nu_l})^2 \Big]^2\, .
\end{equation}
Combining this last theoretical prediction with the limit in Eq.\ 
(\ref{rhoexp}), we find for $m=100$~GeV ($m=0$)
\begin{equation}
  \label{rhoMF}
\bigg(\frac{M_F}{1\ {\rm TeV}}\bigg)\ 
\stackrel{>}{{}_\sim}\ \left\{ \begin{array}{l}  
2.2\ (25)\times \bigg(\sum\limits_{l=e,\mu,\tau} h^2_l\bigg)\,,
\quad {\rm for}\ \delta=2 \\ \\
0.4\, d^{1/2}_\delta\, \bigg(\sum\limits_{l=e,\mu,\tau} h^2_l\bigg)\
\frac{\displaystyle S_\delta}{\displaystyle \delta - 2}\ ,\quad 
{\rm for}\ \delta > 2
\end{array}\ \right.
\end{equation}
The above constraints  are much weaker than  those derived from limits
on non-oblique new-physics observables.

Before  closing Section 5, we wish  to comment on  the fact that there
may exist theoretical  uncertainties related to  the one-loop results. 
In the loop calculations, it is usually  considered that the KK theory
is  truncated  at some   scale  $M'_F$ close to   $M_F$,  where $M'_F$
represents the  energy scale  of the  active  KK states,  beyond which
string effects are  still   negligible \cite{DDG0}.  In   the  present
analysis, we assumed that $M'_F =  M_F$.  Depending on the dynamics of
a given  string theory, however,  one  generally has  $M'_F \le  M_F$,
i.e.\  $x_F  \equiv M'_F/M_F  \le 1$.  Thus,  for $x_F   \not= 1$, the
non-decoupling contributions discussed in  Section 4 (e.g.\ the $Zll'$
coupling given   by  Eq.\ (\ref{TZll2}))  must   be multiplied  by  an
additional  factor $(x_F)^{2\delta}$. For  example, if $x_F = 0.9$ and
$\delta = 6$, this additional multiplicative factor is $\sim 0.3$.  On
the  other hand,  there may  exist compensating  factors, which can be
obtained     by   modestly   rescaling  the     normalization  of  the
higher-dimensional  Yukawa couplings  $h_l$  in  Eq.\ (\ref{hl}).  For
instance, if $h_l$ is rescaled by a factor 1.35 for $\delta = 6$, then
the factor $(x_F)^{2\delta}$, with $x_F = 0.9$,  drops out completely. 
Because of the  above  theoretical uncertainties, which  are generally
inherent in all truncated  KK theories, the numerical predictions  for
the  new-physics phenomena and the  derived   limits should rather  be
viewed as order-of-magnitude estimates.

\setcounter{equation}{0}
\section{Conclusions}

We studied the phenomenological implications of Kaluza--Klein theories
of low-scale quantum gravity, which may naturally predict, in addition
to  gravitons, the presence  of singlet neutrinos.  In these theories,
the singlet neutrinos can  propagate  in a  higher-dimensional  space,
which  is endowed  with a  number  $\delta$  of  large compact spatial
dimensions.     Such   low-scale    KK     theories    that    include
higher-dimensional singlet  neutrinos  are very appealing,  since they
can naturally  provide suppression   mechanisms for  understanding the
smallness in  mass of  the observed  light  neutrinos \cite{ADDM,DDG},
explain the observed anomalies in solar  and atmospheric neutrino data
through neutrino oscillations  \cite{DDG,DS}, and  finally account for
the baryon asymmetry in the Universe through leptogenesis \cite{AP}.

Another very interesting feature  of KK theories of  low-scale quantum
gravity is that the presence of KK-neutrino states  can give rise to a
number of testable new-physics signals at collider and lower energies,
such  as lepton-flavour violation in muon,   tau and $Z$-boson decays,
coherent $\mu\to e$ conversion  in nuclei, and effects of universality
breaking  in  $Z$-  and $W$-boson  interactions.    We confronted  the
predictions of a minimal higher-dimensional singlet-neutrino model for
the aforementioned  new-physics phenomena  with  current  experimental
data.  We  found that  KK neutrinos heavier   than the  $W$  boson act
cumulatively  in  the  loops,   thereby  leading to  a  non-decoupling
phenomenon, for large values of the original higher-dimensional Yukawa
couplings of  the theory.   Because of this  cumulative non-decoupling
phenomenon of the KK  states, we  were  able to derive  very stringent
constraints on lepton-flavour violation in the $\mu$-$e$ sector, which
originate from upper limits  on $\mu \to e$  conversion in nuclei  and
$\mu \not\!  \to eee$.  In fact, the limits derived in this way can be
much stronger than  those obtained from  $\mu \not\!  \to e \gamma$ by
an earlier   consideration  \cite{FP}, for  higher-dimensional  Yukawa
couplings of order unity.   Specifically, for $h_e =  h_\mu = h_\tau =
1$,     we   derived  the    rather     tight lower     limits:  $M_F\
\stackrel{>}{{}_\sim}\ 380\ (1430)$ and 310 {\rm  TeV}, for 2 and 3--6
large extra dimensions, respectively, and $m=100$~GeV ($m=0$).  As was
illustrated  at the end  of Section 5  however, we should stress again
that these lower limits are quite generic  and must only be considered
as  order-of-magnitude   estimates   due  to    inherent   theoretical
uncertainties  of       the  string dynamics       at  the fundamental
quantum-gravity scale $M_F$.

New-physics effects in $e$-$\tau$  and  $\mu$-$\tau$ sectors are  less
constrained than those  found for the  $e$-$\mu$ one.  In general, one
may avoid    most   of  the   latter    constraints  by  setting   the
higher-dimensional  Yukawa  couplings $h_e$    or  $h_\mu$ to   zero.  
However, even    if  $h_e$ and  $h_\mu$ are    vanishingly  small, the
non-observation of universality breaking in leptonic $Z$-boson decays,
which    involve  $\tau$ leptons in  the    final state, together with
experimental  data on the invisible $Z$-boson  width are sufficient to
place a lower bound on $M_F$. For $h_\tau = 1$  and $h_\mu = h_e = 0$,
the lower limit on $M_F$ was found to be: $M_F\ \stackrel{>}{{}_\sim}\ 
3$  TeV, almost independently     of   the number of   large   compact
dimensions.   Even though  the analysis  was  performed  in a  minimal
higher-dimensional  model, these  last    bounds, as  well    as those
pertaining  to     the $e$-$\mu$    sector,   are  generically  rather
model-independent, as long as order-unity Yukawa couplings, $h_l$, are
present in the original    theory before compactification.    In  this
context, it is important  to  emphasize again  that all of  the  above
`one-loop' lower limits on  $M_F$ scale quadratically with $h_l$,  for
$h_l\stackrel{>}{{}_\sim} 1$. This  is an important consequence of the
cumulative non-decoupling  effect  that  results from summing    up KK
states heavier than the electroweak scale.  For instance, if $h_l \sim
3$, the  lower bounds  on $M_F$ will   then increase  by one  order of
magnitude.  On the other hand, for  maximal values of $h_l\sim 5$, the
branching ratios for  $\tau \to e\mu\mu$  and $Z\to e\tau$ could reach
the $10^{-7}$  and  $10^{-6}$  levels, respectively.   Next-generation
colliders have the potential  capabilities to probe  these predictions
of KK  theories   of low-scale quantum   gravity  and  to   impose new
constraints on their parameters.

\subsection*{Acknowledgements} 
We  wish  to   thank  Nima  Arkani-Hamed  and  Savas    Dimopoulos for
discussions.  AP thanks the  Theory Groups  of  SLAC and Fermilab  for
their kind hospitality, while part of this work was done.

\end{document}